%% file: main.tex
\documentclass[journal]{IEEEtran}

\ifCLASSINFOpdf

\else

\fi
\usepackage{nccmath,amssymb}
\usepackage[nocompress]{cite}

\usepackage{amsmath}
\usepackage{colortbl}
\usepackage{mathtools}

\usepackage[colorlinks=true,linkcolor=red,breaklinks=True,citecolor=red]{hyperref}
\PassOptionsToPackage{hyphens}{url}

\PassOptionsToPackage{dvipsnames}{xcolor}

\usepackage{enumitem}
\setlist[itemize]{noitemsep, topsep=0pt}

\usepackage{adjustbox}
\usepackage[linesnumbered,algo2e,ruled]{algorithm2e}
\usepackage{multirow}
\usepackage{booktabs,subcaption,dcolumn}
\usepackage{tikz}
\usepackage{enumitem}
\usepackage{tabularx}
\usepackage[font=footnotesize]{caption}

\usepackage{pifont}
\usepackage{svg}
\usepackage{soul}

\newcommand{\smallsim}{\smallsym{\mathrel}{\sim}}

\makeatletter
\newcommand{\smallsym}[2]{#1{\mathpalette\make@small@sym{#2}}}
\newcommand{\make@small@sym}[2]{%
  \vcenter{\hbox{$\m@th\downgrade@style#1#2$}}%
}
\newcommand{\downgrade@style}[1]{%
  \ifx#1\displaystyle\scriptstyle\else
    \ifx#1\textstyle\scriptstyle\else
      \scriptscriptstyle
  \fi\fi
}
\makeatother

\def\figurename{Fig.}

\let\OLDthebibliography\thebibliography
\renewcommand\thebibliography[1]{
  \OLDthebibliography{#1}
  \setlength{\parskip}{0pt}
  \setlength{\itemsep}{0pt plus 0.3ex}
}

\usepackage{amsmath}
\usepackage{amsfonts, amssymb}
\usepackage{colortbl}

\newcommand{\cmark}{\color{green!60!black}{\ding{51}}}
\newcommand{\xmark}{\color{red!80!black}{\ding{55}}}

\newcommand{\tm}[1]{\fontfamily{ppl}{ #1}}%
\newcolumntype{?}{!{\vrule width 1.5pt}}

\newcommand{\takeaway}[1]{
    \vspace{1em}
    \noindent\fbox{%
        \parbox{\columnwidth}{%
            \textbf{Takeaway}. {#1}
        }%
    }
}

\newcommand{\threatmodel}[1]{
    \vspace{1em}
    \noindent\fbox{%
        \parbox{\columnwidth}{%
            {#1}
        }%
    }
}

 \newcommand*\whitecirc[1][0.7ex]{%
  \begin{tikzpicture}
  \draw (0,0) circle (.10);
  \end{tikzpicture}}
  
   \newcommand*\blackcirc[1][0.7ex]{%
  \begin{tikzpicture}
  \fill (0,0) circle (.10);
  \end{tikzpicture}}

\newcommand{\ballnumber}[1]{\tikz[baseline=(myanchor.base)]
\node[circle,draw,fill=white,inner sep=1pt] (myanchor) {\color{black}\bfseries\footnotesize #1};}

\usepackage[outline]{contour}

\newcommand{\csctu}[1]{CS1}
\newcommand{\cselasticmon}[1]{CS2}
\newcommand{\csirish}[1]{CS3}
\newcommand{\csrml}[1]{CS4}
\newcommand{\cspamimo}[1]{CS5}
\newcommand{\csdeepslice}[1]{CS6}

\newcommand{\overbar}[1]{\mkern 1.5mu\overline{\mkern-1.5mu#1\mkern-1.5mu}\mkern 1.5mu}

\newcommand\dataset[1]{{\fontfamily{pcr}\selectfont {\footnotesize #1}}}

\AtBeginDocument{%
  \providecommand\BibTeX{{%
    \normalfont B\kern-0.5em{\scshape i\kern-0.25em b}\kern-0.8em\TeX}}}

\def\BibTeX{{\rm B\kern-.05em{\sc i\kern-.025em b}\kern-.08em
    T\kern-.1667em\lower.7ex\hbox{E}\kern-.125emX}}
\begin{document}

\title{Wild Networks: Exposure of 5G Network Infrastructures to Adversarial Examples}

\author{
\IEEEauthorblockN{Giovanni Apruzzese, Rodion Vladimirov, Aliya Tastemirova, Pavel Laskov\\}
\IEEEauthorblockA{\textit{Institute of Information Systems - University of Liechtenstein}\\
\{name.surname\}@uni.li}\\
}

\markboth{IEEE Transactions on Network and Service Management}%
{Shell \MakeLowercase{\textit{et al.}}: Bare Demo of IEEEtran.cls for IEEE Journals}

\maketitle

\input{sections/0-abstract.tex}


\IEEEpeerreviewmaketitle

\input{sections/1-introduction}
\input{sections/2-background}

\input{sections/3-model_new}

\input{sections/4-framework_new}

\input{sections/5-evaluation}

\input{sections/6-related}
\input{sections/7-conclusions}

\bibliographystyle{IEEEtran}

\input{main.bbl}

\input{biographies/bio}

\appendices

\input{sections/A1-details}

\end{document}

%% file: sections/0-abstract.tex
\begin{abstract}
Fifth Generation (5G) networks must support billions of heterogeneous devices while guaranteeing optimal Quality of Service (QoS). Such requirements are impossible to meet with human effort alone, and Machine Learning (ML) represents a core asset in 5G. ML, however, is known to be vulnerable to adversarial examples; moreover, as our paper will show, the 5G context is exposed to a yet another type of adversarial ML attacks that cannot be formalized with existing threat models. Proactive assessment of such risks is also challenging due to the lack of ML-powered 5G equipment available for adversarial ML research.

To tackle these problems, we propose a novel adversarial ML threat model that is particularly suited to 5G scenarios, and is agnostic to the precise function solved by ML. In contrast to existing ML threat models, our attacks do not require any compromise of the target 5G system while still being viable due to the QoS guarantees and the open nature of 5G networks. Furthermore, we propose an original framework for realistic ML security assessments based on public data. We proactively evaluate our threat model on 6 applications of ML envisioned in 5G. Our attacks affect both the training and the inference stages, can degrade the performance of state-of-the-art ML systems, and have a lower entry barrier than previous attacks. 
\end{abstract}

%% file: sections/1-introduction.tex
\section{Introduction}
\label{sec:1-intro}

\IEEEPARstart{T}{he} Fifth Generation (5G) network technology standard represents a revolutionary paradigm in the context of telecommunications. On the one hand, it must provide connectivity to billions of heterogeneous devices. On the other hand, it must ensure content delivery offered by thousands of vendors and assure excellent Quality of Service (QoS)~\cite{morocho2019machine}.
Such requirements--despite bringing undeniable benefits to the end-users of 5G--represent a serious problem for the \textit{tenants} of the 5G Network Infrastructures (NI). The available resources (e.g., bandwidth or battery capacity) in the 5G NI are limited, and sophisticated management is required to meet the 5G standards. 
Hence, orchestrating the 5G NI demands timely and precise response to changing environments, and exclusive reliance on hand-crafted or hardcoded methods is impractical.
To solve this problem, many works suggest~\cite{5Garchitecture, morocho2019machine}, or endorse~\cite{5GMLrec}, the integration of Machine Learning (ML) in the 5G NI. 
For instance, \textit{network slicing}---emblematic of 5G---is focused on dynamic resource allocation and can greatly leverage the automation of ML~\cite{kafle2018consideration}.
Nonetheless, ML automation can only be appreciated with the `standalone' (SA) implementation of 5G~\cite{gsma:2020}, whose deployment has just begun~\cite{Verizon:5GSA}.\footnote{Most operational 5G NI leverage the `non-standalone' (NSA) architecture, a mere an enhancement of the previous 4G NI.}
However, to ensure reliable future telecommunication systems, the security of SA 5G must be put under scrutiny, \textit{in advance}~\cite{Biggio:Wild}.

Many studies (e.g.,~\cite{dutta20205g, suomalainen2021securing}) investigated conventional security aspects in 5G.
In contrast, we focus on the specific threat arising from the deployment of ML: \emph{adversarial examples}~\cite{Biggio:Wild}, which can influence the decisions of ML systems via small data manipulations. Some prior works (e.g.,~\cite{flowers2019evaluating, usama2019adversarial, shi2018spectrum}) provide evidence that also ML applications envisioned in 5G are susceptible to adversarial examples.
Despite their high effectiveness, such examples were always generated by attackers conforming to standard ML threat models, e.g., `white-box' or `black-box'. Satisfying the underlying assumptions of all such attacks requires some compromise of the system hosting the target ML `box'---which constitutes a high entry barrier in critical infrastructures~\cite{apruzzese2021modeling}. We observe that the 5G NI is exposed to a much more subtle variant of adversarial examples which cannot be formalized with prior works.

The major contribution of this paper is the proposal of the new \textit{myopic} threat model, which \textit{complements} existing ML threat models. Our threat model highlights that the 5G paradigm enables attacks that can be launched from the adversary's legitimately owned devices, without compromising any segment of the 5G NI.
In contrast to previous work (e.g.,~\cite{davaslioglu2019trojan, sadeghi2019physical, usama2021examining, kim2020over}), our myopic attacker---constrained by the 5G context---has less knowledge and capabilities, but can still damage the 5G NI tenants by exploiting the very foundations of the 5G paradigm, such as its open nature and the QoS guarantees~\cite{komeylian2020deploying, qureshi2020service}. Moreover, our threat model is \textit{agnostic} of the specific ML deployment, allowing coverage of yet to be conceived applications of ML in 5G.

The novelty of our threat model and the nascent rollout of SA 5G demand the respective ML security evaluations that follow the \emph{proactive} approach endorsed by Biggio and Roli~\cite{Biggio:Wild}:  ``identifying relevant threats... and simulating the corresponding attacks; devising suitable countermeasures; and repeating this process \emph{before}\footnote{Emphasized by Biggio and Roli.} system deployment''. However, \textit{any} scientific effort on 5G ML security must face a tough challenge: no real and ML-powered 5G system is currently available for adversarial ML research, and attacking the real 5G NI is not ethical. On the other hand, defeating the ML system on in-house data is not scientifically convincing due to the risk of experimental bias. 
To solve this dilemma, we propose a \textit{generic framework} for security evaluations of 5G ML components based on open-source data validated by the research community. In this framework, we explicitly define the data transformations yielding realizable adversarial perturbations~\cite{tong2019improving} that target ML components of the 5G NI.

Using the proposed framework, we proactively evaluate our threat model through 6 case studies, epitomizing different ML functions in the 5G NI envisioned by the state-of-the-art. We assess myopic attacks at inference and training stage, and also gauge some existing countermeasures. Our findings show that myopic attacks degrade the performance of state-of-the-art ML systems for the 5G NI. Despite being less effective than prior white-/black-box attacks, myopic attacks have a lower entry barrier and can still impact the 5G NI tenants. 

\textbf{Contribution and Organization.} In summary, this paper makes the following contributions to the state-of-the-art:
\begin{itemize}
    \item We propose the new \textit{myopic} threat model. This threat model is tailored for attacks against ML systems in 5G NI, and is agnostic of the specific ML application.
    
    \item We present a framework for realistic assessment of adversarial attacks against the 5G NI based on public data.
    
    \item We use our framework to proactively evaluate the myopic threat model against 6 state-of-the-art ML prototypes for diverse 5G tasks. 
\end{itemize}

The paper is structured as follows.
We introduce the 5G NI, the role of ML, and motivate our paper in §\ref{sec:background}.
We present our myopic threat model in §\ref{sec:model}.
We describe our 5G ML security evaluation framework in §\ref{sec:framework}.
We showcase our case studies in §\ref{sec:evaluation}.
We compare our paper with related work in §\ref{sec:related}.
We conclude the paper in §\ref{sec:conclusions}. 

%% file: sections/2-background.tex
\section{Background and Motivation}
\label{sec:background}
This paper spans across three broad areas: 5G Networking, Machine Learning, and Cybersecurity. Understanding all such areas is \textit{fundamental} to understand our contribution.

We begin by presenting an overview of the 5G ecosystem (§\ref{ssec:infrastructure}), and the role of ML in 5G (§\ref{ssec:ML5G}). Then, we delve into ML security aspects (§\ref{ssec:security}). Finally, we highlight the problems of the state-of-the-art that motivate our paper (§\ref{ssec:motivation}). 

\subsection{The 5G Ecosystem}
\label{ssec:infrastructure}
\begin{figure*}[!htbp]
    \centering
    \frame{\includegraphics[width=2\columnwidth]{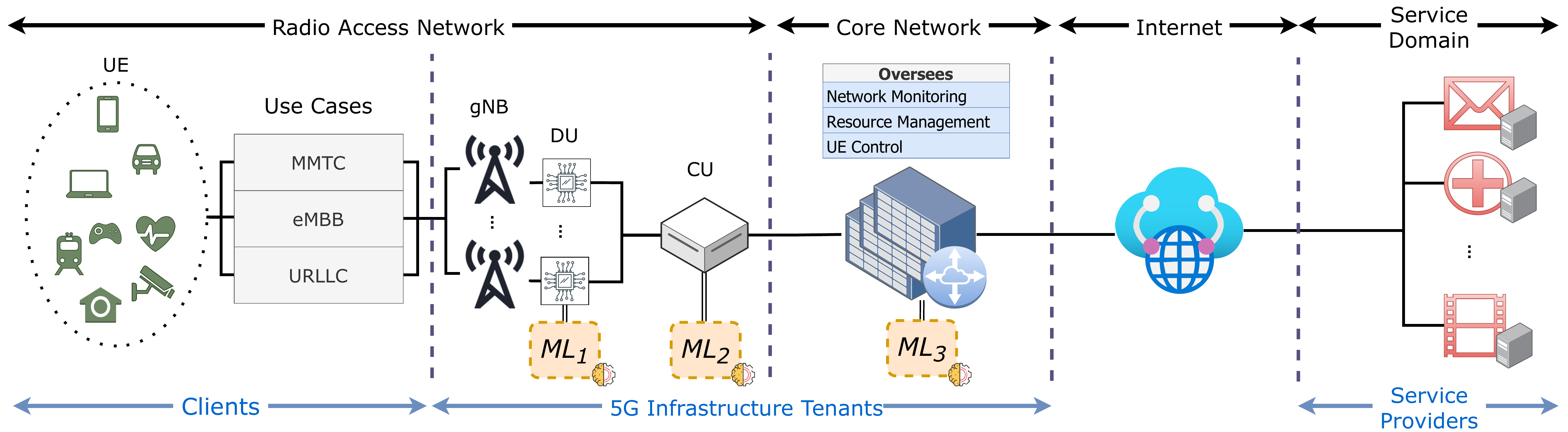}}
    \caption{The 5G Ecosystem. The \textit{clients} transparently use the network infrastructure deployed by the 5G \textit{tenants} to reach the service \textit{providers}.}
    \label{fig:infrastructure}
\end{figure*}

Future generations of mobile networks (5G and beyond) will \textit{by far surpass current mobile communication}. They will certainly provide more capabilities and enhanced experience for human users in form of better bandwidth and lower latency. They will, however, also support billions of other actors such as self-driving cars, autonomous delivery drones, intelligent sensors, wearable medical devices and the like. All these entities must share--and compete for--the limited available resources (e.g., bandwidth, battery power, latency). 
This ``ecosystem'', and the envisioned technical infrastructure behind it, is depicted in Fig.~\ref{fig:infrastructure}.

The 5G Network Infrastructure (NI)
allows users to obtain any remote service, and can be split in two network segments~\cite{5Gspecification}.
The \textit{Radio Access Network} (RAN) leverages the New Radio (NR) standard to provide a physical connection between the user equipment (UE) and the air interface devices in the base stations (gNB) whose main task is to forward user data to the core network\footnote{NSA 5G also uses eNB-ng, pertaining to 4G, and hence outside our scope.}. The gNB integrate specific components, such as distributed units (DU) and centralized units (CU), dedicated to preliminary network management or data preprocessing.
The \textit{core network} forwards user data to the Internet and performs more complex orchestration functions, such as the ``Access and Mobility Management function'' (AMF) which oversees resource optimization and management of UE (e.g., authentication, authorization and billing).

The 5G NI is \textbf{open in nature}~\cite{komeylian2020deploying, 5G:open}, and accessing its functionalities does not require excessive verifications: in most cases, a valid 5G subscription is sufficient for a UE to be included in the 5G ecosystem~\cite{gavrilovska2020cloud, kazemifard2021minimum}. In particular, it is common to group UE into three generic use cases~\cite{5Garchitecture}. The Massive Machine Type Communication (MMTC) is characterized by low data rates and very high connection density. The enhanced Mobile Broadband (eMBB) implies low connection density and high average and peak data rates. The Ultra-Reliable and Low-Latency Communication (URLLC) requires extreme reliability and low latency but assumes low data rates. 

From the economic perspective, three parties interact in the 5G NI: clients, infrastructure tenants, and service providers. Clients request services from service providers. Infrastructure tenants enable service delivery by connecting clients with the service providers. The business relationships between the 5G NI tenants and the other parties are governed by the \textbf{Service Level Agreements} (SLA). Such SLA define the QoS~\cite{qureshi2020service} that must be met by the 5G NI, alongside the penalties for failing to meet such requirements. Compared to the past paradigms in which SLA primarily focused on the overall availability (e.g.,~\cite{mastroeni2011violation}), in 5G the SLA are expected to have a finer granularity tailored to individual clients or services~\cite{accedian:2021, papageorgiou2020sla}.

Managing such heterogeneous ecosystem while ensuring SLA compliance is hardly feasible via static and human-defined methods~\cite{rodrigues2019machine, benzaid2020ai}. To increase the efficiency of 5G networks, \textbf{ML is expected to play a pivotal role in the underlying infrastructure empowering 5G}~\cite{morocho2019machine}. This is a significant difference with respect to previous networking paradigms: in 4G (and older) networks, ML was far from being mature and ready for real deployment.

\takeaway{Three characteristics make the (SA) 5G NI a unique setting compared to other network infrastructures: its openness, the fine-grained SLA, the role of ML. Corporate networks may use ML, and---when outsourced---may have strict SLA, but are not open. Older networks (e.g., 4G) are open, but have more relaxed SLA and do not leverage ML.}

\subsection{Machine Learning for 5G Networking}
\label{ssec:ML5G}
Let us briefly explain why ML represents a valuable asset for sustainable management of the 5G NI.

From the viewpoint of the 5G NI tenants, among the main goals of ML in 5G is \emph{meeting the QoS requirements of the SLA}. 
Failing to reach the agreed levels of QoS violates the SLA and result in penalties for the infrastructure tenants~\cite{papageorgiou2020sla, towersexchange:2019}. 

ML techniques can be deployed by the infrastructure tenants anywhere in their 5G NI (indicated as multiple $ML$ boxes in Fig.~\ref{fig:infrastructure}).
Let us showcase four functions envisioned in 5G for which ML proved to be successful by the state-of-the-art.

\textit{Network slicing}~\cite{5Gspecification} aims at dividing network resources to optimize the delivery of specific groups of services. 
For example, entertainment video streaming requires high bandwidth but can tolerate temporary packet losses. In contrast, loss of information in eHealth services is unacceptable, but they do not require high bandwidth to operate correctly. Slices may correspond to the three 5G use-cases (eMBB, URLLC and MMTC); a ML approach for assigning UE to such slices was proposed in~\cite{thantharate2019deepslice}; but other criteria are also possible~\cite{kafle2018consideration,li2018deep, coronado2019flow}, including using ML for application-based slices~\cite{le2018sdn, le2018applying}. With respect to Fig.~\ref{fig:infrastructure}, an exemplary ML-based module for network slicing may correspond to the ${ML}_{3}$ box. 

\textit{Automatic Modulation Recognition} (AMR) aims to infer modulation type of a given signal~\cite{o2016radio}. While AMR was initially proposed for military purposes, it is now an established function of 5G networking~\cite{hermawan2020cnn} and can greatly benefit from ML~\cite{hazar2018performance}. By identifying the correct modulation, it is possible to guarantee optimal transmission quality by tuning the transmission parameters accordingly. In Fig.~\ref{fig:infrastructure}, a ML module for AMR can be deployed in the ${ML}_{1}$ box.

Another function of 5G networking is \textit{predicting the Channel Quality Indicator} (CQI). The CQI measures the communication quality received by UE: a UE computes and reports the CQI to its serving gNB; the gNB uses such CQI to ensure optimal transmissions. Communicating the CQI from the UE to the gNB has a high overhead, and frequent mutations in the environment may change the CQI before it is sent to the gNB. To avoid QoS degradation, ML can be used to predict the CQI, either via past reportings~\cite{kimura2021deep, parera2019transfer}, or by using other metrics communicated more frequently to the gNB and correlated to the CQI~\cite{vasilakos2020integrated, ul2020supervised}. In Fig.~\ref{fig:infrastructure}, an exemplary ML module for CQI prediction can be deployed in the ${ML}_{1}$ box.

Recently, the utility of deep learning has been demonstrated for \textit{power allocation} in massive Multiple Input-Multiple Output (mMIMO) scenarios~\cite{van2020power}, a cornerstone of 5G. The idea of mMIMO is to deploy a large number of antennas at gNB, which transmit the same data to UE. The transmission capacity is measured by \emph{spectral efficiency} (SE), which depends on the power allocated to each UE served by a gNB. Such allocation can be computed with ML by analyzing the position of the UEs in a given environment~\cite{sanguinetti2018deep}. 
In Fig.~\ref{fig:infrastructure}, a ML module for power allocation can be placed in the ${ML}_{2}$ box.

These four functions will be the target of our demonstration in §\ref{sec:evaluation}.
The set of functions that exploit ML in 5G is much broader~\cite{morocho2019machine}, and some are still to be conceived.

\subsection{Security of Machine Learning}
\label{ssec:security}
Security analysis require the notion of a \textit{threat model} describing the viewpoint of the attacker w.r.t. the target system:
depending on the \textit{knowledge} and \textit{capability}, the attacker follows a specific \textit{strategy} to reach the intended \textit{goal}~\cite{Biggio:Wild}.

By focusing on ML security, the so-called `adversarial attacks' aim to adversely effect the target ML system via some data perturbation, i.e., \textit{adversarial examples}~\cite{Biggio:Wild}.
Even imperceptible perturbations (e.g., a single pixel~\cite{Su:One}, or few extra bytes~\cite{Apruzzese:Evaluating}) can compromise the decisions of ML systems.

Threat models focused on ML security hence revolve around such adversarial examples. An adversary may have a targeted or an indiscriminate \textit{goal}~\cite{barreno2010security}, e.g., affecting specific examples, or degrading the overall performance of the ML system.
An attacker may have variable degrees of \textit{knowledge} on three key ML elements~\cite{Laskov:Practical}: the ML model, $M$; the training data, $T$; and the feature set, $F$. On the other hand, the \textit{capability} defines how the attacker can interact---possibly under some external constraints~\cite{Pierazzi:Intriguing}---with such elements: e.g., manipulate $T$, affect $F$, or use $M$ as an `oracle'~\cite{Papernot:Practical}. Finally, the attack \textit{strategy} depends on the previous assumptions: e.g., introducing `backdoors' in $T$~\cite{quiring2020backdooring}, or creating a surrogate $M$ and transferring the successful examples to the real $M$~\cite{Demontis:Adversarial}.
 
Classical ML attack scenarios are often expressed via the notion of a 'box'.
A \textit{white-box} attacker has full knowledge of $M$ and $F$ enabling the generation of optimal perturbations~\cite{flowers2019evaluating}. 
In a \textit{black-box} setting~\cite{Dang:Evading}, the attacker has no knowledge of $M$, $T$ or $F$, but can query $M$ (possibly subject to some query budget~\cite{ilyas2018black}). In gray-box settings, the attacker may have some knowledge for optimizing the querying strategy~\cite{apruzzese2020deep}. In \textit{no-box} settings, the attacker cannot interact with the system, but knows $T$~\cite{Chen:Zoo}. A further distinction is made according the the attacker's impact on $T$. If the attacker has write-access (direct or indirect) to $T$, then she can affect the \textit{training} stage of $M$~\cite{Biggio:Poisoning}, otherwise she is limited to attacks at \textit{inference} stage~\cite{Biggio:Evasion}. 

Despite an explosive interest to security of ML no universal countermeasures against adversarial examples have been identified so far. Some approaches (e.g., \textit{adversarial training}~\cite{tramer2018ensemble} or \textit{feature removal}~\cite{smutz2012malicious}) require foreseeing the exact form of adversarial examples. Other techniques are applicable only when the entire feature space is completely modifiable by the attacker, or to image-related data (e.g., \textit{certified defenses}~\cite{lecuyer2019certified}). 
Regardless, \textit{any} defense may degrade performance in the absence of adversarial attacks~\cite{he2017adversarial}.

\subsection{Motivation, Challenges and Scope}
\label{ssec:motivation}

Due to its relevance in SA 5G, ML represents an attractive attack target, and adversarial examples require a dedicated treatment w.r.t. traditional security analyses~\cite{Biggio:Wild}. However, related researches in 5G ML are immature, in particular: (i) \textit{existing ML threat models are inadequate}; (ii) \textit{realistic security assessments are difficult}. We aim to overcome both of these shortcomings which are now discussed in more detail. 

The assumptions of classical adversarial scenarios are hard to meet in security sensitive environments. In the 5G NI context, they essentially imply a full security breach. For instance, having complete knowledge of a \textit{trained} ML model for a white-box attack requires that an adversary gains access to a respective 5G NI component, which is well protected by conventional security mechanisms. The same holds true for black-box attacks since the output of ML models may only be visible from within the 5G NI. The training data for such critical components is likewise well protected and cannot be freely manipulated. Although meeting the assumptions of classical ML threat model is in principle possible, compromising the 5G NI is difficult and costly,  and attackers may opt for different strategies. Due to the open nature of 5G, an attacker in possession of UE has legitimate access to the 5G RAN. We hence focus on how such an attacker can leverage adversarial examples to inflict damage to the 5G NI tenants by targeting their ML systems. Such offensive strategies are not covered by the state-of-the-art and require formalizing specific threat model---which must be proactively evaluated for deployments of ML in high-risk scenarios~\cite{Biggio:Wild, Europe:AIreg}.

Realistic assessments of adversarial examples require reproduction of the physical constraints binding the attacker ~\cite{Pierazzi:Intriguing, tong2019improving}. This is particularly challenging in the 5G context:
due to the early stage of SA 5G~\cite{Verizon:5GSA}, all such systems are protected by NDA\footnote{We interviewed multiple telcos which confirmed this fact.} and no ML-equipment is currently available for research. Furthermore, there is a lack of publicly available data for reproducing state-of-the-art ML systems~\cite{bonati2020open}. 
Finally, practical evaluations must also consider countermeasures and their potential degradation to the baseline performance.

To overcome these shortcomings, we propose the novel \textit{myopic} threat model (§\ref{sec:model}) which is complementary to existing ML threat models and is specifically tailored to the 5G paradigm.
We also present a new framework for security assessment of 5G ML (§\ref{sec:framework}), which explains how to leverage existing public data to craft realizable adversarial perturbations against state-of-the-art ML components envisioned in 5G. Finally, we apply our framework to evaluate our threat model via 6 case studies (§\ref{sec:evaluation}) considering different adversarial scenarios.

This work is focused on attacks against ML in SA 5G. Security issues not related to ML, or pertaining to different ML applications, are beyond the scope of this paper. We stress that we are not the first to consider adversarial ML attacks against the 5G NI. We directly compare our work with the state-of-the-art at the end of this paper (§\ref{sec:related}) because the differences of our efforts w.r.t. previous research can only be appreciated after thoroughly understanding our major contributions.

%% file: sections/3-model_new.tex
\section{Myopic Threat Model}
\label{sec:model}
Our primary contribution is the ``myopic threat model'', describing adversarial ML attacks against the 5G NI which are feasible to stage due to their low cost. Indeed, the corresponding adversarial examples can be generated from the UE of an end-user (i.e. the attacker), without any access to the 5G NI. Before explaining how this is possible, let us introduce the notation for adversarial ML attacks.

Let $M$ be a (trained) ML model that analyzes a set of features $F$ that describe some data samples; let $x$ be a given sample, let $F_x$ be the feature representation of $x$. Assume that $M$ can correctly predict the ground truth of $F_x$ as $M(F_x)$. In an adversarial attack, a small perturbation $\epsilon$ causes $M$ to predict a wrong output on the input $F_x$.
We distinguish two scenarios: (i) \textit{Feature-space Perturbations} (FsP), if $\epsilon$ is applied to the feature representation of $x$, thus resulting in ${F_x + \epsilon}$; or (ii) \textit{Problem-space Perturbations} (PsP), if $\epsilon$ is applied via the process that generates $x$, thus resulting in $x+\epsilon$. The attack aims to finding such perturbation $\epsilon$ and can be formalized as: 
\begin{equation}
\resizebox{0.6\columnwidth}{!}{
  {{find $\epsilon$ s.t. }}$
    \begin{cases}
      M(F_x + \epsilon) \neq M(F_x) & \text{\small{FsP}}\\
      M(F_{x + \epsilon}) \neq M(F_{x}) & \text{\small{PsP}}\\
    \end{cases}$
    }
    \label{eq:AA}
\end{equation}

It is implicitly assumed that $F_{x+\epsilon}$ and ${F}_{x}+\epsilon$ are associated with the same ground truth as $F_x$. 
Also note that the problem- and feature-space can overlap if $F_{x+\epsilon}\!=\!x+\epsilon$, meaning that FsP and PsP can be equivalent~\cite{Pierazzi:Intriguing}.

\subsection{Definition of the Myopic Threat Model}
\label{ssec:myopic}

\begin{figure}[!htbp]
    \centering
    \includegraphics[width=1\columnwidth]{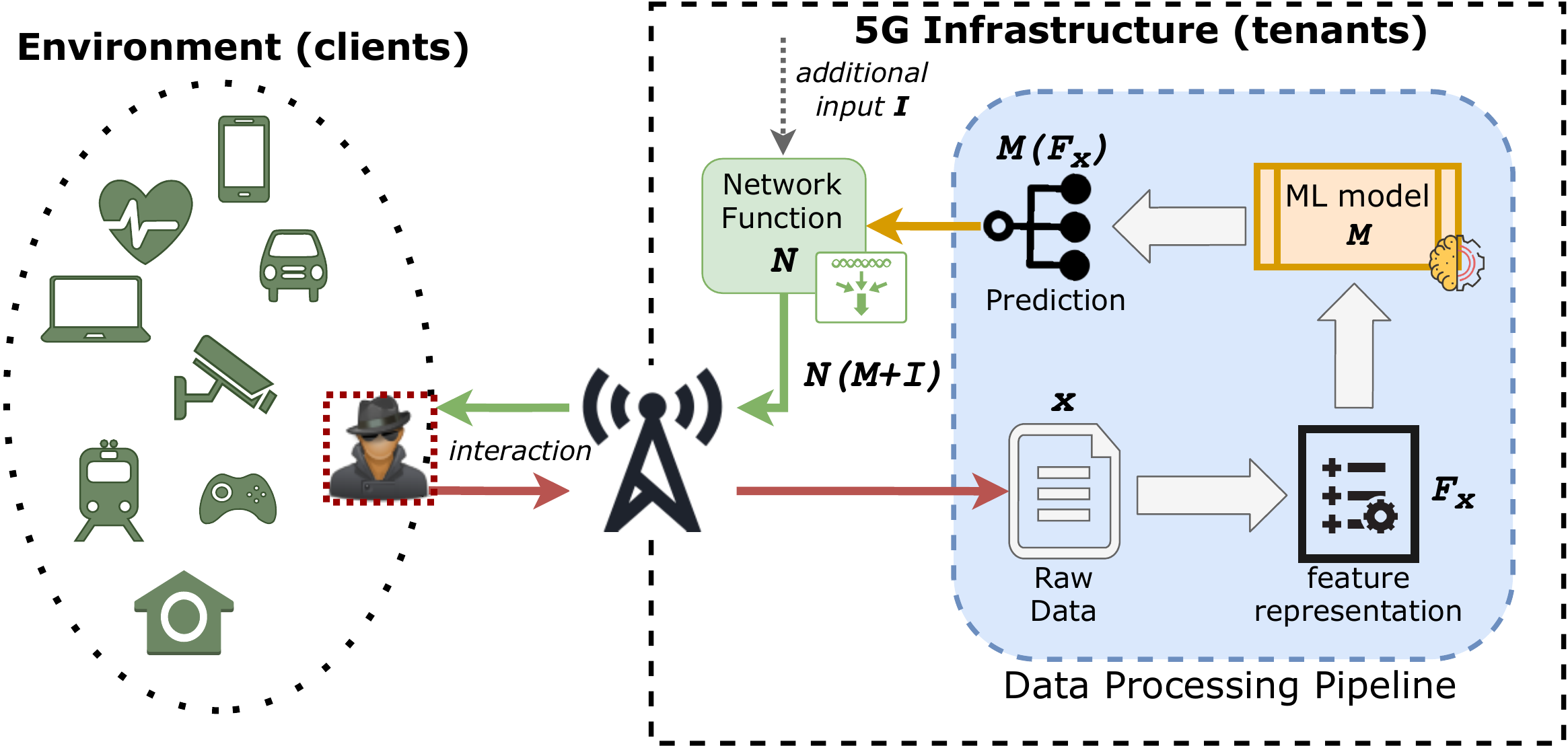}
    \caption{Illustration of our myopic threat model.} 
    \label{fig:scenario}
\end{figure}

To substantiate the claim that our threat model is realistic, let us interpret the abstract setting described by Eq.~\ref{eq:AA} using the general characteristics of 5G networking. This scenario is depicted in Fig.~\ref{fig:scenario}.
The attacker operates as a client within a 5G NI and hence is part of a heterogeneous environment with many UEs. The attacker uses her UE to interact with the 5G NI (owned by its \textit{tenants}). The results of such interactions `enter' the 5G NI in the form of raw-data $x$, which is subject to some preprocessing. These operations yield $F_x$, which is passed as input to a ML model, $M$, that contributes to a given network function, $N$ (e.g., network slicing). Hence, the prediction of the ML model, $M(F_x)$, is sent to $N$. However, besides $M(F_x)$, the network function $N$ may use \emph{additional input}, $I$, that does not depend on $x$, e.g., actions from other UEs, or state of the 5G NI.
The output of the network function, $N(M\!+\!I)$, is not directly visible from outside the 5G NI, but all the UEs in the environment (including the attacker's) may be affected by it. 
For instance, the resources allocated via network slicing depend on, and affect, the entire environment.

These assumptions describe a \emph{myopic} attacker, whose `sight' does not go beyond her UE. 

We can now formally define the myopic threat model according to the four criteria described in §\ref{ssec:security}:

\threatmodel{
\tm{\textbf{Goal.} } 
The attacker intends to \textit{inflict damage to the 5G NI tenants} via adversarial perturbations against the ML model $M$ used by the network function $N$.

\tm{\textbf{Knowledge.} }
\textit{The attacker has limited knowledge of the target system.} She only knows a subset of features ${\mathcal{F}} \subseteq F$ used by $M$. The attacker does not know the exact implementation of either $N$ or $M$, and can observe neither $M(F_x)$ nor $I$, and not even $N(M+I)$.
Finally, the attacker has no information about the training set $T$ used to develop $M$.

\tm{\textbf{Capability.} }
\textit{The attacker has no control on the 5G NI, but has full control of her UE.}
She can freely manipulate the behaviour of her UE to influence her interactions with the 5G NI and thus can consciously affect a subset of features $\overbar{\mathcal{F}} \subseteq \mathcal{F}$ via PsP.
The attacker cannot modify the raw-data $x$ after it is acquired by the 5G NI and has no access to the data processing pipeline. 

\tm{\textbf{Strategy.} }
The attacker \textit{guesses} a PsP affecting some features within her knowledge and capabilities, $\overbar{\mathcal{F}}$, analyzed by $M$.
}

\noindent
Let us make three important remarks.
\begin{itemize}
    \item When the PsP is translated into the \textit{feature space}, such PsP will affect the features $\overbar{F} \! \supseteq \! \overbar{\mathcal{F}}$; i.e., the PsP may \emph{inadvertently} affect features beyond the attacker's knowledge.
    
    \item Our threat model is \textit{agnostic} of the task solved by $M$ and is applicable to a broad range of 5G use-cases.
    
    \item The PsP can be generated by manipulating the UE's behaviour \emph{at any layer of the protocol stack}.  For instance, an attacker may use some app in a different way, modify an open-source app to affect the network layer, or manipulate the regular functionality of the UE's operating system or its hardware (e.g., the well-known \textit{flashing}~\cite{zhang2014once}).\footnote{Such changes, specially at lower layers, \emph{may only possible} via modifying the respective functionality of the UE.} 
\end{itemize}
In our demonstration (§\ref{sec:evaluation}) we will investigate six ML case-studies, highlighting: (i) the roles\footnote{To the best of our knowledge, we are the first to clearly distinguish these four feature sets. Such complexity is necessary to simulate all the imperceptible effects that can occur within the 5G NI.} played by $F$, $\overbar{F}$, $\mathcal{F}$, $\overbar{\mathcal{F}}$; and (ii) the effects of PsP on such sets.\footnote{In §\ref{sec:framework} we propose an equivalent transformation to PsP that can be leveraged by research endeavours to replicate physically realizable perturbations.}

\subsection{Comparison with Previous Threat Models}
\label{ssec:comparison}

Let us elucidate how the myopic threat model is \textit{complementary} to the conventional ML threat models; Table~\ref{tab:models} summarizes this discussion. The key difference lies in the semantics of the `box' considered by the respective attack scenarios. 

\textbf{Our `box'.} In the myopic threat model, the `box' is the entire 5G NI (i.e., the dotted square in Fig.~\ref{fig:scenario}). This is a complex system made of thousands of components to which the attacker has no internal access. The attacker's UE interact only with the 5G RAN. The received feedback is the response of the 5G NI to the entire environment, and hence too complex to be usable. Furthermore, the attacker cannot directly interact with the ML component $M$. Due to such limited knowledge, essentially restricted to $\mathcal{F}$ and further aggravated by the PsP contraint, the attacker is confined to a rough guessing strategy.

\textbf{Conventional `box'.} In contrast, in traditional ML threat models the `box' is the ML component itself (e.g., the orange box in Fig.~\ref{fig:scenario}). Hence, the attacker can directly observe the impact of her actions to the predictions of the ML component (as in gray/black-box scenarios). Moreover, having complete knowledge of $M$ (in white-box attacks), or of both $F$ and $T$ (for no-box attacks), enables creation of ML components that are an exact match of the real $M$. In all such scenarios, the attacker can leverage her knowledge and capabilities to optimize her perturbations. Even the `physical' attacks introduced in~\cite{kurakin2017adversarial} differ from the scenario envisioned in our threat model: they account for the separation between physical input and its data representation; however, the prediction of $M$ corresponds to the output of the `box'. In contrast, our threat model also separates the ML's predictions from the feedback received by an attacker \textit{by two layers of indirection}: the network function $N(M+I)$, and the final feedback of the 5G NI---which is not immediately usable by our attacker due to its complexity.

\begin{table}[!htbp]
    \centering
    \caption{Myopic threat model vs existing `box' threat models.}
    \label{tab:models}
    \resizebox{0.95\columnwidth}{!}{
        \begin{tabular}{c?c|c|c|c|c}
             & 
             \rotatebox{0}{ \begin{tabular}{c} \textbf{White} \\ \textbf{box} \end{tabular}} &
             \rotatebox{0}{ \begin{tabular}{c} \textbf{Gray} \\ \textbf{box} \end{tabular}} &
             \rotatebox{0}{ \begin{tabular}{c} \textbf{Black} \\ \textbf{box} \end{tabular}} &
             \rotatebox{0}{ \begin{tabular}{c} \textbf{No box} \end{tabular}}  & 
             \rotatebox{0}{\cellcolor{gray!10} \begin{tabular}{c} \textit{\textbf{Myopic}} \end{tabular}} \\
            \toprule
            \begin{tabular}{c} \textbf{Available} \\ \textbf{Knowledge} \end{tabular} & $M, F$ & $\mathcal{F}$ & \xmark & $F, T$ & \cellcolor{gray!10} $\mathcal{F}$ \\ \midrule
            
            \begin{tabular}{c} \textbf{Optimal} \\ \textbf{Perturb.} \end{tabular} & \cmark & \cmark & \cmark & \cmark & \cellcolor{gray!10} \xmark  \\ \midrule
            \begin{tabular}{c} \textbf{ML prediction} \\ \textbf{$M(F_x)$} \end{tabular} & \cmark & \cmark & \cmark  & \xmark & \cellcolor{gray!10} \xmark \\ 
             
            \bottomrule
        \end{tabular}
    }
\end{table}

Finally, we note that the myopic attacker is \textit{less} powerful---and hence \textit{more} realistic---than attackers typically assumed in cryptoanalysis settings (irrespective of the existence of ML). For instance, the Dolev-Yao threat model~\cite{zhang2019formal} assumes attackers with complete control of both ends of the communication channel (i.e., the UE and the 5G NI); in contrast, our myopic attacker only controls one end (i.e., her own UE).

\subsection{Viability of Myopic Attacks in 5G}
\label{ssec:viability}

Let us explain why the `guessing' strategy of the myopic attacker is particularly viable in 5G. To this end, we must connect our threat model (§\ref{ssec:myopic}) with the unique characteristics of the 5G NI (§\ref{ssec:infrastructure}), emphasizing the role played by ML.

We note that real attackers are not interested in crafting the `smallest' perturbation that results in a successful adversarial example\footnote{Crafting the `minimal' perturbation is the typical assumption in adversarial ML literature focusing on computer vision~\cite{Papernot:SoK}.}, and are not bound to any self-imposed `magnitude' constraint~\cite{carlini2019robustness}. Indeed, real attackers operate with a cost/benefit mindset~\cite{wilson2014some}: if their goal is reached at an `affordable' cost, then any strategy is viable. In the case of a myopic attacker, such \textit{goal} is ``cause damage to the 5G NI tenants by targeting ML with adversarial examples.''

Malfunctioning of ML in 5G is likely to cause QoS degradation in the environment (e.g., poor connection) or damage the 5G NI equipment (e.g., battery depletion). Since QoS is tied to SLA in the 5G ecosystem, any QoS degradation---even that of the attacker's own UE---may be a reason for filing a complaint with the respective regulatory authority, leading to financial damages for the 5G NI tenants~\cite{sciancalepore2017mobile}. The open nature of the 5G NI further aggravates the problem: the attacker can legitimately introduce a large number of UEs into the environment and thus amplify the attack impact at low cost.  

The combination of binding SLAs and the openness of the 5G NI makes myopic attacks both feasible and harmful\footnote{We interviewed several telcos, which acknowledged such risk.}, especially given the lower entry barrier compared to other threats to 5G. To execute a myopic attack, no compromise of the UE or the NI components is needed. 

Finally, the assumption that the attacker has full control of the UE is also well-founded and typical in security analyses. Some manipulations may require bypassing the basic security mechanisms of an UE; however, the attacker must learn how to break such mechanisms only \textit{once} for all her UEs (assuming the same brand).\footnote{All the attacks in our demonstration (§\ref{sec:evaluation}) leverage well-known data manipulation techniques.}

\takeaway{
Existing ML threat models can be invalidated by denying an attacker internal access to the 5G NI. In contrast, a myopic attacker can inflict damage to the 5G NI tenants by merely changing the behavior of her UE. Such novel threat requires a dedicated treatment and proactive evaluation.}

%% file: sections/4-framework_new.tex
\section{5G ML Security Evaluation Framework}
\label{sec:framework}

To set up the stage for security evaluations of ML in 5G, we present a \textit{high-level  framework} focused on \textit{fair} and \textit{realistic} assessments of adversarial ML threat models\footnote{Our framework complements those that do not assume ML (e.g.,~\cite{hussain2018lteinspector}).}. 

Various previous works investigated ML methods that can empower the 5G NI (cf.~§\ref{ssec:ML5G}). However, reproducing such methods on a in-house and closed environment--and showing the effectiveness of ad-hoc adversarial perturbations--introduces a substantial experimental bias. 
To mitigate such bias and enable a \textit{fair} assessment, our framework is based on open-source data. As a main contribution, our framework ensures the evaluation of realistic adversarial ML scenarios in 5G---a challenging task, given the current state-of-the-art.

\subsection{Suitable Public Data}
\label{ssec:requirements}
Using publicly available data for security evaluations requires such data to meet four criteria. Specifically, a given public dataset $X$ is \textit{suitable} if: (i) it is validated by the research community, i.e., created by adopting state-of-the-art methodologie; (ii) it contains the ground truth information; (iii) it complies with 5G; and (iv) it allows creation of realistic adversarial examples. Let us elucidate the last two criteria.

\textbf{Compliance with 5G.} Despite the existence of several tools to perform 5G simulations~\cite{bonati2020open}, only few state-of-the-art works release (e.g.,~\cite{sanguinetti2018deep}), or are fully built on (e.g.,~\cite{thantharate2020secure5g}), public data.   
Most prior works on 5G ML do not disclose any dataset (e.g.,~\cite{coronado2019flow, li2018deep}), preventing accurate replication of their solutions. Robustness of such ML systems \emph{can be assessed on other data} provided that such data relates to the same 5G task. This may require to \textit{adapt} a given dataset, e.g., by removing some samples or features. Such intuition can lead to discovery of `novel' datasets usable for 5G ML evaluations---despite the fact that they were released when 5G was not yet defined or before its rollout began.

\textbf{Realistic Perturbations.} Relying on pre-collected data clearly makes it impossible to operate in the problem space. This challenge impairs the simulation of realistic adversarial examples which assume physically realizable perturbations~\cite{tong2019improving}. Our solution to this challenge is to leverage perturbations applied in the \textit{raw-data space}, which can be made semantically equivalent to PsP. This technique is presented in the following section.

\subsection{Raw-data Space Perturbations}
\label{ssec:rsp}

Let us illustrate our intuition by recalling Fig.~\ref{fig:scenario}. Here, we can see that the 5G NI `receives' a raw-data sample, $x$, which is translated into its feature representation, $F_x$, and then forwarded to the ML model, $M$. Such operations are invisible to the ML model.
Hence, if a perturbation $\epsilon$ is applied directly to $x$, then its effects carry over the entire preprocessing pipeline until $F_{x+\epsilon}$ is created and analyzed by $M$. Through such \textit{Raw-data space Perturbations} (RsP) it is possible to retroactively simulate PsP \emph{in a research environment} by using existing datasets.  

To be semantically equivalent to a PsP, the RsP must consider the influence of a real attacker on the data generation process and anticipate the effects of such influence on the corresponding raw-data. Specifically, the following criteria must be met:
\begin{itemize}
    \item The perturbation must reflect the \textit{capabilities} assumed in the threat model. For instance, some values simply cannot be influenced by an attacker.
    \item After its application, the \textit{integrity} of the perturbed raw-data must be preserved. For instance, some values depend on others, and such dependencies must be updated. 
    \item The perturbation must abide by the \textit{constraints} of the data generation process. For instance, the payload of a TCP packet must have between 0 and 1500 bytes. 
\end{itemize}
An RsP that meets these criteria is equivalent to a PsP and hence suitable for realistic security assessments. Otherwise, such RsP either violates the assumptions of the threat model or results in a `trivial' adversarial example that would be rejected by the data processing pipeline before reaching $M$.

Finally, depending on the specific use-cases, an RsP may have different \textit{intensity} as long as the underlying constraints are met. For instance, a perturbation may entail ``sending some extra bytes'', but the amount of such extra bytes can vary (e.g.,~\cite{Apruzzese:Evaluating}). 

\takeaway{To evaluate realistic attacks via public data, the `mapping' between the data generation process and the actual data contained in the dataset must be taken into account. This requires: a dataset, $X$, containing raw-data for the 5G NI (cf. $x$ in Fig.~\ref{fig:scenario}); and anticipating a real attacker's effects on such raw-data by means of an RsP.}

\subsection{Workflow}
\label{ssec:workflow}

The proposed framework is aimed at \textit{any} research on ML security in the 5G context. Hence it is agnostic of the specific purpose and functionality of the ML component, the format of the dataset, and of the threat model itself. 

Our framework has three inputs: a dataset $X$ meeting the criteria in §\ref{ssec:requirements}; the specifics to devise a state-of-the-art ML model $M$ with $X$; and the details to create RsP on $X$.

\begin{figure}[!htbp]
    \centering
    \includegraphics[width=0.95\columnwidth]{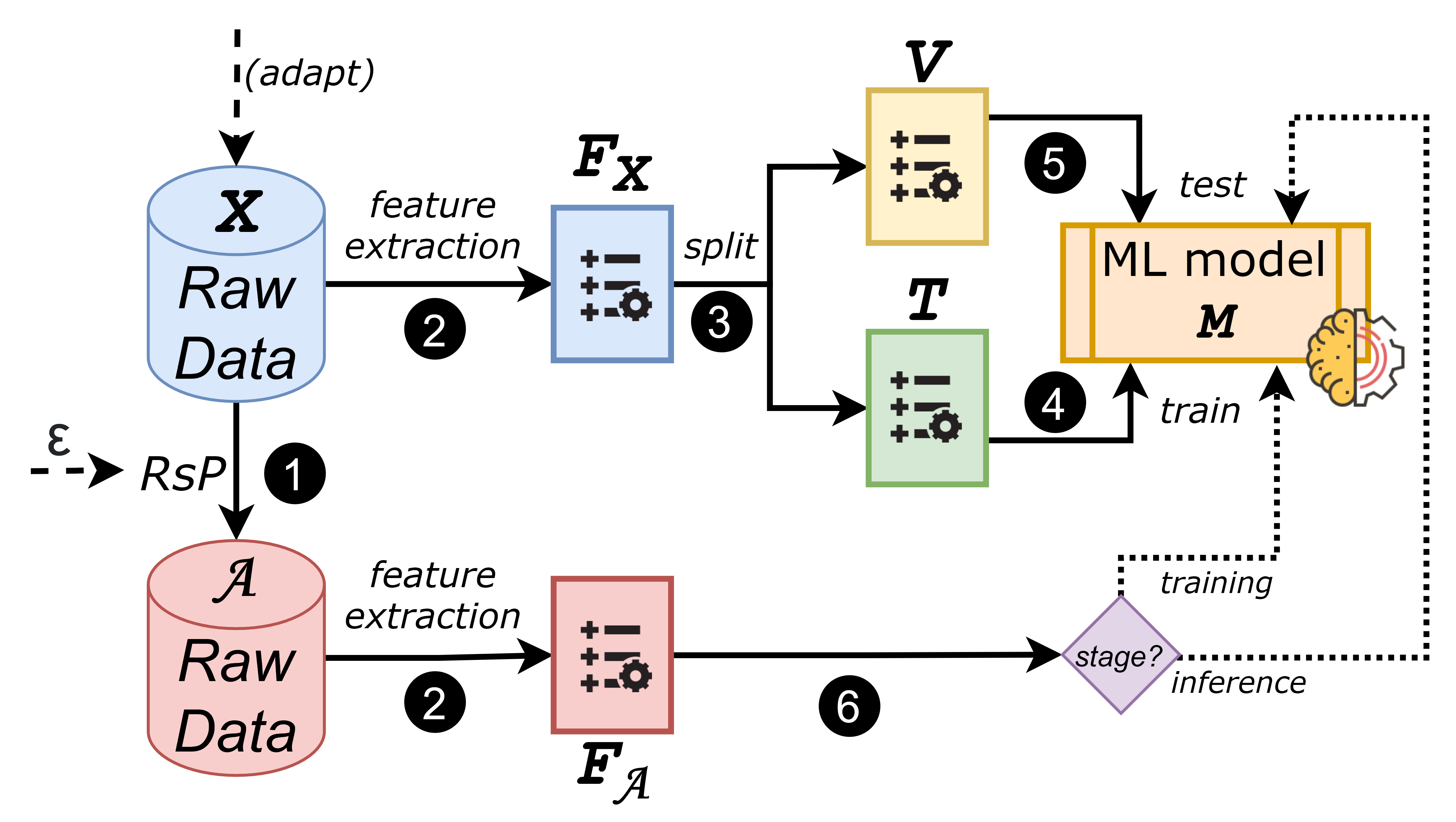}
    \caption{Workflow of the proposed 5G ML security evaluation framework.}
    \label{fig:framework}
\end{figure}

We provide a high-level schematic of our framework in Fig.~\ref{fig:framework}. 
After acquiring a given dataset $X$ (and adapting $X$ to comply with 5G, if required), the first step \ballnumber{1} is applying any sensible RsP on a pre-defined subset of $X$, yielding the adversarial subset of raw-data $\mathcal{A}$. Such RsP must meet all the criteria listed in §\ref{ssec:rsp}, and further verifications may be necessary.
Then, \ballnumber{2} both $X$ and $\mathcal{A}$ are subject to the \textit{same} feature extraction operations, resulting in $F_{X}$ and $F_{\mathcal{A}}$ respectively.
Next, \ballnumber{3} $F_{X}$ is split into a training $T$ and a validation $V$ partition\footnote{We consider $T$ and $V$ to be in their feature representation.}, used to train \ballnumber{4} a ML model $M$ and assess \ballnumber{5} its baseline performance.  
Finally, \ballnumber{6} the adversarial examples in $F_{\mathcal{A}}$ are used to `attack' $M$. To simulate \textit{inference} stage attacks, the performance of $M$ is assessed on $F_{\mathcal{A}}$; otherwise, for attacks at \textit{training} stage, $M$ is re-trained using both $T$ and $F_{\mathcal{A}}$, and its performance is assessed again on $V$.

Proactive security evaluations must also consider \textit{countermeasures}. As stated in §\ref{ssec:security}, existing defenses may have limited applicability, and can induce \textit{performance degradation} in the absence of adversarial attacks. Nobody would be interested in a defense that protects against an (uncommon) attack at the expense of an unusable ML system in (routine) operation. The workflow in §\ref{ssec:workflow} can account for the effects of a countermeasure by repeating the same steps for a `hardened' version of $M$, which we denote as $\hat{M}$. We define the \textit{tradeoff} of a countermeasure on $M$ as: 
\begin{equation}
    \label{eq:tradeoff}
    {\mathbb{T}(M)} = {P}_{M} / {P}_{\hat{M}}
\end{equation}
\noindent
where $P_M$ (resp. $P_{\hat{M}}$) is the performance of $M$ (resp. $\hat{M}$) on the same validation data $V$, according to a given performance metric $P$ (e.g., Accuracy, F1-score). 
An effective defense should mitigate an attack while achieving $\mathbb{T}$ close to $1$.

%% file: sections/5-evaluation.tex
\section{Proactive Evaluation of Myopic Attacks}
\label{sec:evaluation}
As a proactive demonstration, we apply the proposed evaluation framework to assess myopic attacks against ML components for 5G NI. Our intention is to showcase myopic attacks in a broad array of ML deployment scenarios in SA 5G---to the extent this is possible given the current state-of-the-art. To the best of our knowledge, our paper provides the largest evaluation of adversarial attacks against ML systems envisioned in the 5G NI. 

\textbf{Overview.}
Our evaluation spans over 6 Case Studies (CS), each having a different scope. Specifically:
\begin{itemize}
    \item \csctu\ ~(§\ref{ssec:ctu}): we assume a myopic attacker controlling \textit{multiple} UEs, and assess the impact of adversarial examples at both training and inference stages; moreover, in this CS all the feature sets ($F$, $\overbar{F}$, $\mathcal{F}$, $\overbar{\mathcal{F}}$) are disjoint.
    
    \item \cselasticmon\ ~(§\ref{ssec:elasticmon}): we compare the efficacy of two well-known defenses for a ML \textit{regressor}; we also vary the \textit{strength} of the attacker by considering different $\overbar{\mathcal{F}}$.
    
    \item \csirish\ ~(§\ref{ssec:irish}): we consider attacks against an \textit{online} ML system that leverages time-series analysis.
    
    \item \csrml\ ~(§\ref{ssec:rml}): we compare the robustness of \textit{shallow} and \textit{deep learning} against a myopic attacker; we also compare the effectiveness of myopic attacks against \textit{white/black-box} attacks performed by past work against the exactly same ML system. 
    
    \item \cspamimo\ ~(§\ref{ssec:pamimo}): we consider a \textit{single-} and \textit{multi-agent} attack scenario targeting a \textit{physical} quality metric.
    
    \item \csdeepslice\ ~(§\ref{ssec:deepslice}): we showcase a ML system that is \textit{secure-by-design} against myopic attacks.

\end{itemize}
Each CS considers a different public dataset, used to reproduce state-of-the-art ML systems for a networking task of 5G NI. We provide an aggregated discussion in §\ref{ssec:discussion}. Extensive technical details of our CS are provided in Appendix~\ref{app:testbed}, which also includes an overview of the chosen datasets (Table~\ref{tab:datasets}).

\textbf{Evaluation Procedure.}
Each CS represents a unique setup, but the evaluation follows the same procedures, all based on the workflow of §\ref{ssec:workflow}.
First, we elucidate the considered 5G system by summarizing the \textit{data-stream} that pertains to the targeted ML component. Then, we use a public dataset to train a ML model $M$ using a given set of features $F$ and assess the performance of $M$.
We then simulate myopic attacks against $M$. We explicitly describe the \textit{knowledge} of the attacker by specifying $\mathcal{F}$, the subset of features the attacker is aware of. We then elucidate the attacker's \textit{capabilities} by specifying $\overbar{\mathcal{F}}$, the subset of features she can consciously influence. Next, we apply some RsP to the raw-data that affects $\overbar{\mathcal{F}}$ (and hence $\overbar{F} \supseteq \overbar{\mathcal{F}}$).  
Finally, we verify the integrity of the perturbed raw-data to ensure that all dependencies are preserved and the reported values are correct. 

Note that a myopic attacker cannot craft an optimal adversarial example and can only guess a desired perturbation (cf. §\ref{ssec:viability}). To account for a broad range of potential perturbations, we consider an array of RsP targeting each feature in $\overbar{\mathcal{F}}$ at different \textit{intensity}.\footnote{Due to the novelty of our threat model, we will consider defenses that are `generally' applicable. For instance, we will not consider \textit{certified defenses} because they are tailored for adversarial attacks on images and because they assume attackers that are bound by a given perturbation magnitude---both of which are assumptions that do not pertain to our threat model.}

In our CS, we will attack both `baseline' ML systems, as well as `stronger' ML systems that integrate some defensive mechanism. Because our focus is on ML security, we consider countermeasures against adversarial examples---which is the typical approach in adversarial ML literature (e.g.,~\cite{carlini2021poisoning, apruzzese2020deep}). Some CS are expanded with comparisons with white/black-box attacks; or with considerations on some protection strategies that do not belong to the ML domain.

\input{sections/CS/ctu}
\input{sections/CS/elasticmon}

\input{sections/CS/irish}
\input{sections/CS/rml}

\input{sections/CS/pamimo}

\input{sections/CS/deepslice}

\input{sections/discussion}

%% file: sections/CS/ctu.tex
\subsection{\csctu\ : Network Slicing}
\label{ssec:ctu}
\textbf{Highlights.} The myopic attacker owns \textit{multiple} UEs, and affects both the \textit{inference} and \textit{training} stages. We also assess \textit{defensive distillation} as a countermeasure. Finally, $F$, $\overbar{F}$, $\mathcal{F}$ and $\mathcal{\overbar{F}}$ are all distinct.

\textbf{Target 5G system.}
According to the state-of-the-art (e.g.,~\cite{coronado2019flow, li2018deep, le2018applying}), ML can support 5G network slicing by analyzing Network Flows (NetFlows). NetFlows capture metadata about the communication sessions in a given network. In this CS, ML is used to distinguish \textit{active} (e.g., web-browsing) from \textit{passive} (e.g., an automated update-check) communications. The intuition is that UE involved in active communications should be assigned to slices with higher importance.

The data-stream of this CS can be modeled as follows. The UEs of the entire environment communicate (in the form of network packets) with the 5G NI, which forwards such data to the service domain. Hence, all the raw-packets passing through the 5G NI are captured by the 5G NI and then exported to NetFlows, which are sent to a dedicated ML component that predicts whether such NetFlows belong to passive or active communications. Such predictions are then further elaborated within the 5G NI to apply the respective slicing policies. The ML component is trained on trusted NetFlows, and is periodically updated with new data to prevent performance degradation due to distribution shifts~\cite{sagduyu2019adversarial}.

\textbf{Dataset and Baseline.}
For this CS, we use the \dataset{CTU13}~\cite{Garcia:CTU} dataset, containing multiple traces of \textit{real} network traffic. 
The data in \dataset{CTU13} comes in the form of raw packet captures (PCAP), enabling the application of RsP\footnote{The \dataset{CTU13} also contains the processed NetFlows, which we do not consider because it is not raw-data, and hence not valid for RsP.}. The creators of \dataset{CTU13} provide information allowing to distinguish active (e.g., a human user behaviour) from background (e.g., some passive or scheduled tasks) communications. To the best of our knowledge, we are the first to consider such property of \dataset{CTU13} to attack corresponding ML classifiers through RsP.

To devise the ML system, we use the PCAP traces as basis from which we extract (and label) the corresponding NetFlows by following the exact procedure explained in \dataset{CTU13} documentation. For each PCAP trace we obtain a set of NetFlows, used to devise a Random Forest (\textit{RF}) binary classifier (we devise one classifier per PCAP trace) which analyzes the most common NetFlow fields (we report the complete $F$ in Table~\ref{tab:ctu_features} found in Appendix~\ref{app:CTU}).
Because the majority of \dataset{CTU13} contains  \textit{background} traffic, we use both Accuracy (\textit{Acc}) and F1-score (\textit{F1}) as performance metrics: \textit{Acc}=$0.99$ and \textit{F1}=$0.81$ (the \textit{F1} focuses on \textit{active} connections).

\textbf{Attacks and Defense.} 
The assumed attacker knows that the target system uses a ML component analyzing NetFlows for slice allocation. NetFlows summarize communications between two endpoints with unique \textit{IP} addresses, such as the \textit{duration} (\textit{Dur}), the \textit{ports}, or the \textit{packets} (\textit{Pkt}) and \textit{bytes} (\textit{Byt}) exchanged. The attacker infers such information, hence ${\mathcal{F}}$=(\textit{IPs,Dur,Ports,Pkt,Byt}). However, the attacker knows that she can only influence a subset of ${\mathcal{F}}$: the \textit{IP} is assigned by the 5G NI, the \textit{Dur} depends on the NetFlow appliance managed by the 5G NI, and the (low) \textit{port} is managed by the service providers; these features are beyond attacker's control. The attacker can only consciously influence $\mathcal{\overbar{F}}$=(\textit{Pkts,Byt}) by sending more packets from her UE or adding junk payloads; doing this, however, will also affect other fields in some incontrollable way. We assume that the attacker owns 6 UE, corresponding to $\smallsim$5\% of the (internal) hosts in \dataset{CTU13}. Through these actions, the attacker can affect the target ML system both at \textit{inference} stage (e.g., an `active' communication that gets assigned to a `background' slice), and at \textit{training} stage (therefore inducing poisoning attacks). The latter is due to the well-known fact that ML systems must be continuously updated to prevent concept-drift~\cite{imran2014challenges}, meaning that some RsP may be inadvertedly included in the training data used to update the ML component.

To craft RsP, we extract the raw traffic of the 6 UE owned by the attacker from the raw PCAP traces and perturb the sent packets. Doing so will induce modifications in the \textit{Pkts} and \textit{Byt} features that fall within $\overbar{\mathcal{F}}$, but also other features are affected when the raw packets are transformed into NetFlows. Specifically, $\overbar{F}$=(\textit{Dur},\textit{SrcByt},\textit{DstByt}), because $F$ distinguishes between source and destination bytes, and some packets will be included in other NetFlows; moreover, in the case of TCP connections such actions will also elicit minimal changes to the responses of the contacted host. 

For \textit{inference-stage} attacks, we submit the myopic NetFlows to the baseline \textit{RF}: our objective is assessing the performance degradation of the 6 UEs owned by the attacker. For \textit{training-stage} attacks, we inject the myopic NetFlows into the training set by randomly replacing the original NetFlows of the 6 myopic UEs with their myopic variants; we stress that such procedure does not violate our assumptions because the myopic UEs are trusted by the 5G NI (otherwise, such UEs would not receive any connectivity).
We \textit{do not} manipulate any sample generated by a non-myopic UE (which correspond to 95\% of \dataset{CTU13}). We then re-train the \textit{RF} on such `poisoned' dataset and re-evaluate it on the original testing partition: our objective, here, is assessing the impact on the entire environment. The process is repeated for increasing replacement ratios (25\%, 50\%, 75\%, 90\%) of myopic NetFlows, which correspond to just 1--5\% of the training data.

To investigate a countermeasure against our attacks, we apply the variant of the \textit{defensive distillation} technique suitable for \textit{RF}, as reported in~\cite{Apruzzese:Hardening}. The tradeoff $\mathbb{T}$ of this countermeasure in the absence of attacks, as measured by the Accuracy, is $0.97$, showing a slight increase in the baseline performance. We provide more technical details in Appendix~\ref{app:CTU}, including the detailed workflow for the RsP.

\begin{figure}[!htbp]
    \centering
    \includegraphics[width=\columnwidth]{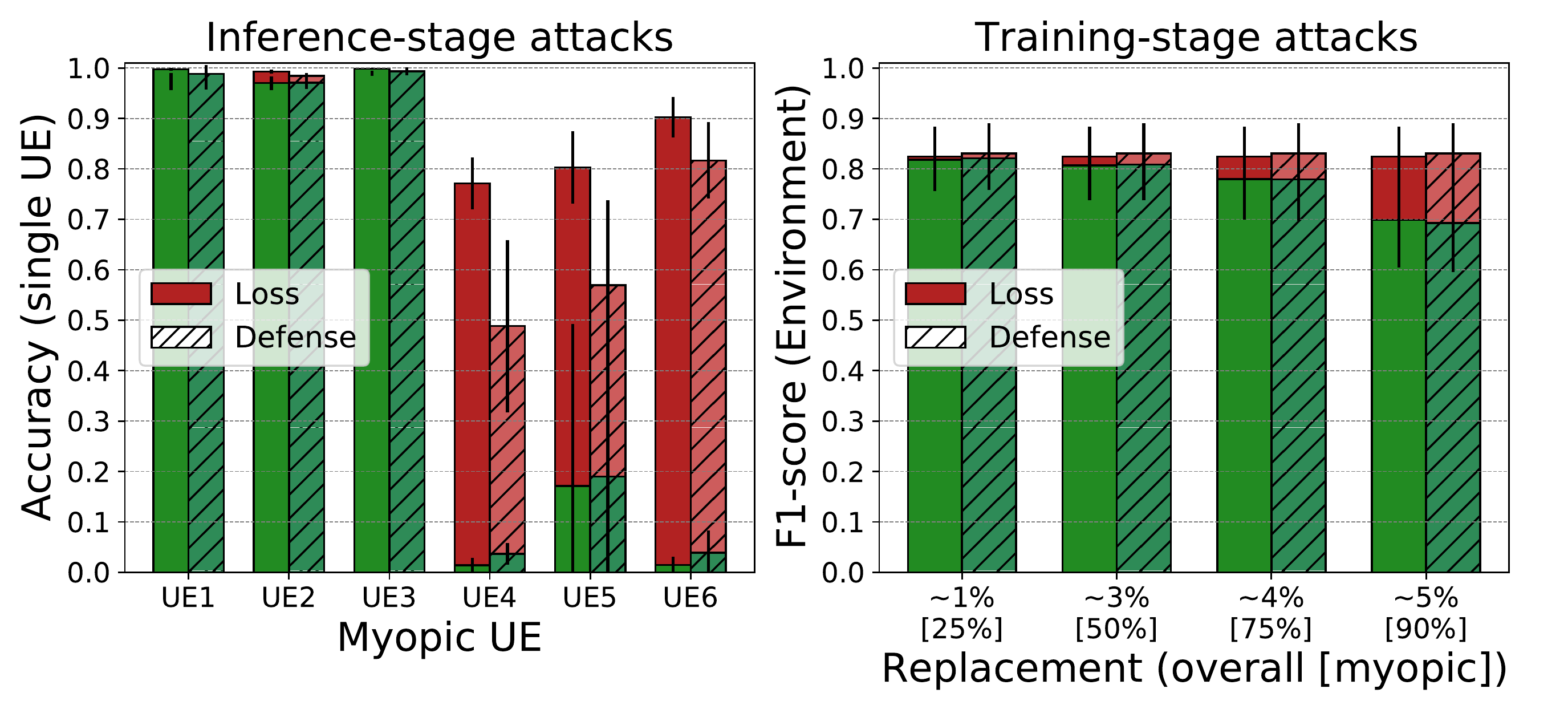}
    \caption{\csctu\ : Myopic attacks at inference (left) and training (right) stage. The metrics are shown for the baseline \textit{RF} (solid bars) and its hardened counterpart (bars with oblique lines). In both graphs, the red part shows the performance degradation induced by the attack; the black line shows the standard deviation over the multiple considered PCAP traces. Inference stage attacks focus on the effects on each myopic UE (denoted by each pair of bars). Training stage attacks focus on the entire Environment, and the performance is reported by varying replacements of NetFlows \textbf{only of} the myopic UE (corresponding at most to 5\% of the overall traffic of the Environment). }
    \label{fig:results_ctu}
\end{figure}

\textbf{Results.}
We report the results of our attacks in Fig.~\ref{fig:results_ctu}.

Attacks at inference stage have notably different effects on the myopic UEs. Some UEs are only slightly affected (UE1--UE3), but others are strongly affected (UE4--UE6). This interesting phenomenon showcases the limited capability of the myopic attacker, who cannot craft optimal and successful perturbations: some UE may be a `useless' vector of RsP, but others can be very powerful---a result which can be amplified by introducing more UEs in the environment.
While these attacks only influence the predictions for the attacker's UEs, we observe that such predictions are used as basis for resource allocation, and hence the entire network can be adversely affected by these attacks.

Attacks at training stage start to impact the whole network after at least 50\% of the attacker's UEs NetFlows are replaced with their myopic variants. Recall, however, the attacker's UEs represent at most $5\%$ of the entire network population, hence such impact is not negligible. 

Defensive distillation is not effective against these attacks; however, it slightly improves the baseline accuracy ($\mathbb{T}$=$0.97$). 

Attacking the considered system via white-/black-box attacks is hardly feasible. For example, the attacker must be aware of the \textit{exact} NetFlow exporter (different tools yield different data~\cite{vormayr2020my}), which is confidential information belonging to the 5G NI tenants. At the same time, the attacker cannot leverage the feedback of $M$ because its predictions (passive or active NetFlows) are further elaborated by the 5G NI before applying the slicing policies.

%% file: sections/CS/elasticmon.tex
\subsection{\cselasticmon\ : CQI Prediction}
\label{ssec:elasticmon}
\textbf{Highlights.} We compare the effectiveness of two popular countermeasures, \textit{adversarial training} and \textit{feature removal}, applied to a ML \textit{regressor}. We also consider myopic attackers having different knowledge and capabilities, by varying $\mathcal{\overbar{F}}$.

\textbf{Target 5G system.} This CS considers the 5G task of CQI prediction.
The ML components must infer the CQI based on measurements computed by the gNB or other measurements reported more often (once every ms) by the UE~\cite{ul2020supervised, vasilakos2020integrated}, such as those related to the Radio Resource Control (RRC) protocol. 

The data-stream envisioned in this CS assumes UE that communicate  channel quality metrics to the gNB, which integrates a ML component that analyzes such metrics and estimates the CQI. The data received by the gNB is considered to be trusted, because implementing security mechanisms would increase the overhead and therefore introduce delays that would defeat the entire purpose of using ML. The ML component is trained on data collected by the 5G NI tenants and conforming to diverse types of UE~\cite{vasilakos2020integrated}.

\textbf{Dataset and Baseline.} We use the \dataset{ElasticMon} dataset~\cite{vasilakos2020elasticsdk} to reproduce the state-of-the-art approach in~\cite{vasilakos2020integrated}.
Released in 2020, \dataset{ElasticMon} contains 5G synthetic raw-data denoting the periodic reportings of a UE to its gNB. The creators of \dataset{ElasticMon} considered all the characteristics of 5G in their network environment. To the best of our knowledge, we are the first to consider this dataset in adversarial scenarios.

Our baseline ML model is a \textit{RF} regressor, which we develop by following the exact instructions of~\cite{vasilakos2020integrated}. We use the raw-data in \dataset{ElasticMon} and follow the same preprocessing steps, obtaining their same feature set $F$ (reported in Table~\ref{tab:elasticmon_features} in Appendix~\ref{app:elasticmon}). Our baseline \textit{RF} achieves similar performance as in~\cite{vasilakos2020integrated}, as measured via Root Mean Squared Error (\textit{RMSE}=$0.22$) and accuracy (\textit{Acc}=$0.95$).

\textbf{Attacks and Defense.} 
An attacker aware that the gNB predicts the CQI on the basis of RRC reportings can expect that the ML system analyzes the Resource Signal Reference Power (RSRP); or, alternatively, the transmitted packets or bytes. Hence, $\mathcal{F}$=(RSRP,\textit{Pkt},\textit{Byt}). The \textit{Byt} or \textit{Pkt} can be easily influenced (as explained in §\ref{ssec:ctu}). The RSRP is computed directly by the UE, and the myopic attacker (who physically owns her UE) is fully able to control the reported RSRP (e.g.,~\cite{lichtman20185g}). Indeed, the RSRP measures the strength of a signal received by a UE from all the surrounding gNBs, with the assumption that the UE will connect to the gNB with the best RSRP. However, an attacker can force her UE to connect to a gNB with a suboptimal RSRP, meaning that the 5G NI will receive an RSRP with different value.\footnote{To the best of our knowledge, there is no way for the 5G NI to prevent such occurrence, as the procedure is carried out on the UE.}

Based on $\mathcal{F}$, we consider four attack scenarios: ${\mathcal{\overbar{F}}_1}$=(RSRP); ${\mathcal{\overbar{F}}_2}$=(\textit{Byt}); ${\mathcal{\overbar{F}}_3}$=(\textit{Pkt}); ${\mathcal{\overbar{F}}_4}$=(\textit{Pkt,Byt}). We simulate these scenarios as follows.
For ${\mathcal{\overbar{F}}_1}$, the RSRP is replaced with a randomly chosen RSRP value in \dataset{ElasticMon}. For ${\mathcal{\overbar{F}}_{2-4}}$ we generate RsP at 7 increasing intensity levels, where the \textit{Pkt} (and/or \textit{Byt}) are incrementally increased as a function of their standard deviation across the \dataset{ElasticMon} dataset. We always verify the integrity of the myopic raw-data, from which we obtain their feature representation. All attacks occur at inference-stage.

We consider two well-known countermeasures, \textit{adversarial training} and \textit{feature removal}, requiring to foresee the patterns of adversarial attacks. We apply both countermeasures by assuming correct anticipation of each attack scenario (as in~\cite{tramer2018ensemble, smutz2012malicious}).
Additional details such as $\overbar{F}$ as well as the application of the RsP and the defenses are in Appendix~\ref{app:elasticmon}.

\begin{figure}[!htbp]
    \centering
    \includegraphics[width=0.95\columnwidth]{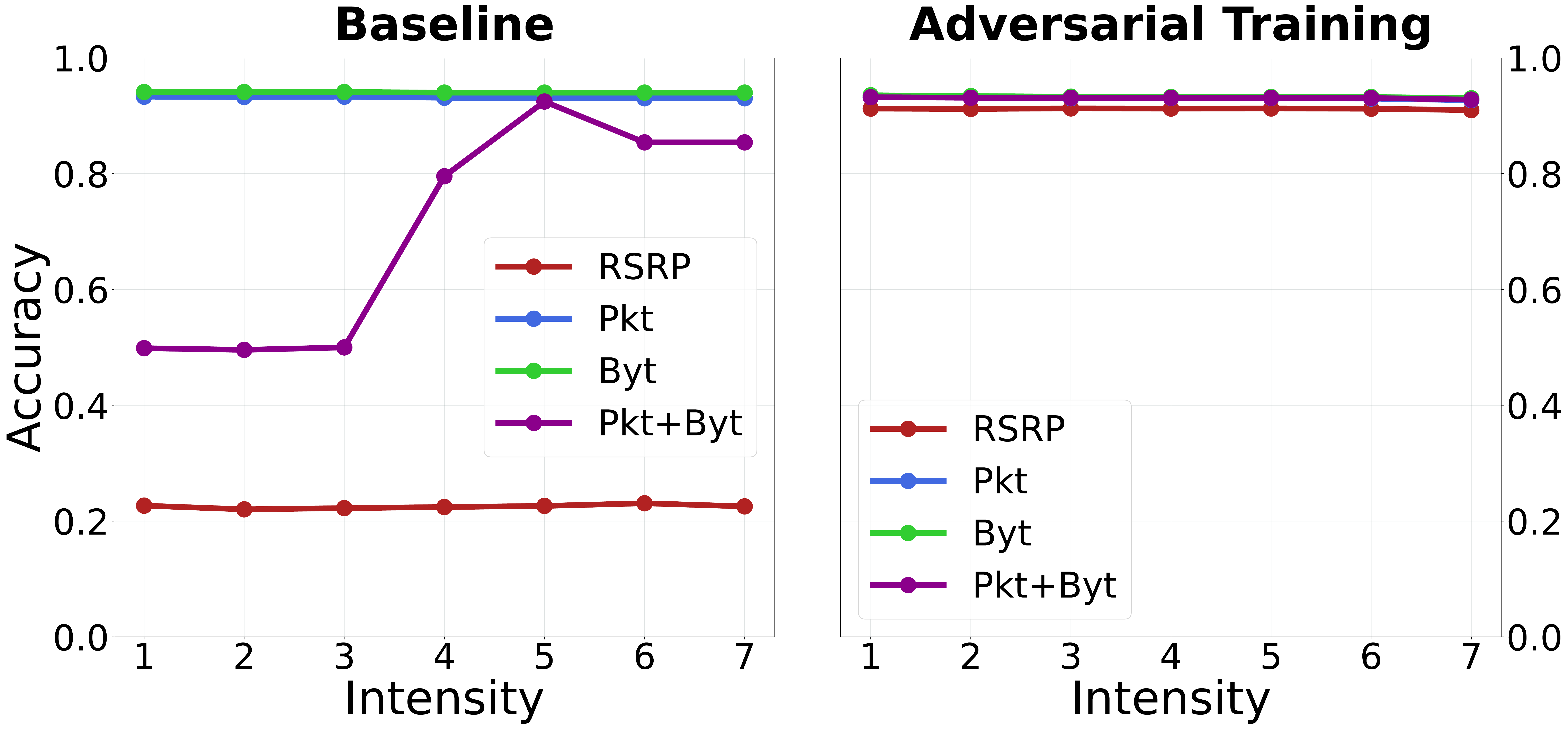}
    \caption{\cselasticmon\ : results for the baseline \textit{RF} (left), and the $\widehat{\textit{RF}}$ hardened through adversarial training (right); feature removal always nullified the attack. The y-axis indicates the accuracy and the x-axis the intensity of each attack (represented with a colored line); for $\overbar{\mathcal{F}_1}$ the results are reported for 7 different random RSRP replacements.}
    \label{fig:results_elasticmon}
\end{figure}

\textbf{Results}.
We report the results of our attacks in Fig.~\ref{fig:results_elasticmon}; we only show adversarial training in Fig.~\ref{fig:results_elasticmon} because applying feature removal \textit{always defused} the attack (no degradation).

By focusing on the baseline \textit{RF} (left graph), we observe that RsP to ${\mathcal{\overbar{F}}_1}$ (red line) lead to significant performance degradation, whereas RsP to ${\mathcal{\overbar{F}}_{2,3}}$ (blue and green lines) have no impact; it is almost counterintuitive that RsP to ${\mathcal{\overbar{F}}_4}$ (purple line) have a higher impact for lower intensities. 
Adversarial training (right graph) protects against all our attacks at little cost (its tradeoff shown in Table~\ref{tab:tradeoff_elasticmon2} is always negligible)---a success shared also by feature removal.

\begin{table}[!htbp]
    \centering
    \caption{\cselasticmon: Tradeoff (lower is better, $\mathbb{T}$=$1$ is no change).}
    \label{tab:tradeoff_elasticmon2}
    \resizebox{0.6\columnwidth}{!}{
        \begin{tabular}{c?c|c|c|c}
             \toprule
             
             \textbf{Defense} & ${\mathcal{\overbar{F}}_1}$ & ${\mathcal{\overbar{F}}_2}$ & ${\mathcal{\overbar{F}}_3}$ & ${\mathcal{\overbar{F}}_4}$ \\
             
             \midrule
             Adv. Tra. & \cellcolor{red!10}$1.01$ & \cellcolor{red!10}$1.01$ & \cellcolor{red!10}$1.01$ & \cellcolor{red!10}$1.01$\\
             Fea. Rem. & $1.00$ & \cellcolor{red!10}$1.01$ & \cellcolor{red!10}$1.01$ & \cellcolor{red!10}$1.01$\\
             
             \bottomrule
        \end{tabular}
    }
\end{table}

Launching adversarial attacks conforming to well-known threat models against the envisioned ML system requires many resources. For instance, obtaining detailed information on $M$ may be prohibitive, because such $M$ is embedded in the gNB, and hence `difficult' to reach---unless the attacker already compromised the 5G NI in some way.

%% file: sections/CS/irish.tex
\subsection{\csirish\ : CQI Prediction (online)}
\label{ssec:irish}

\textbf{Highlight.} We showcase myopic attacks against \textit{online} ML using real 5G network traffic data.

\textbf{Target 5G system.}
This CS also focuses on CQI prediction (as in \cselasticmon{}), but here the prediction is made by analyzing historical CQI reportings via time-series analyses~\cite{kimura2021deep}. The data-stream is similar to the one described in \cselasticmon{}, the only difference being the information communicated by the UE to the gNB (as well as the frequency of such communications).

\textbf{Dataset and Baseline.}
We are inspired by the recent work in~\cite{kimura2021deep} which uses the \dataset{Irish 5G} dataset~\cite{raca2020beyond}. To the best of our knowledge we are the first to evaluate such dataset in adversarial scenarios.
The \dataset{Irish 5G} contains 5G raw-data metrics (including the CQI) collected by the gNB of a major 5G mobile operator and describing 20 minutes of reportings sampled every second. It contains many traces, focused on a different UE mobility pattern, `static' or `driving'. We remove those traces for which the CQI is not provided. The remaining traces refer to activities of the UE, such as `streaming' or `download'. For the `static' mobility pattern, we consider `download' because the CQI of `streaming' never changes; for the `driving' mobility pattern, we consider `streaming' because for the `download' activities the behaviour was too irregular and we never obtained appreciable performance. 

As done in~\cite{kimura2021deep}, our baseline is a Long Short Term Memory (\textit{LSTM}) regressor. Each trace has a dedicated \textit{LSTM}, which consider only the CQI and corresponding timestamp (we do not apply any preprocessing). We use the first half of the trace to pre-train the \textit{LSTM}. Then, the \textit{LSTM} becomes operational and predicts each next CQI value by using the last 30 reportings. The \textit{LSTM} is updated in an online fashion when a new sample is received. Hence $F$=(\text{last 30 CQI}).

\textbf{Attacks.}
A myopic attacker can expect that the 5G NI uses online ML to predict the CQI by using the past history, sampled every second. The attacker cannot know the exact length of such history, hence $\mathcal{F}$=(some past CQI). However, she can be certain that the most recent value is included in such history, hence: $\overbar{\mathcal{F}}$=(previous CQI). The CQI is computed (and reported) by the UE, so the myopic attacker can influence it arbitrarily~\cite{lichtman20185g}. 
Such myopic attacker can affect the ML predictions if the CQI reported by her UE to the gNB is different from the actual one. All subsequent predictions made by considering the history with the myopic CQI will be affected when the ML model will use the `myopic' history to update itself.
We consider two attack scenarios when applying the RsP. In the first scenario, the UE reports that CQI=0, which is the lowest possible value. In the second scenario, the CQI is spoofed with a value within $\pm$3 of the actual one. In both scenarios, the fake CQI is sent once every minute, therefore $\overbar{F}$=(at most one CQI) in both cases. An attacker can theoretically send a fake CQI continuously, but such attempts can hardly be considered as adversarial examples.
Because the \textit{LSTM} are operational for 10 minutes, our RsP will only affect $\smallsim$1\% of the overall sampled data in both scenarios.

\begin{figure}[!htbp]
    \centering
    \includegraphics[width=0.95\columnwidth]{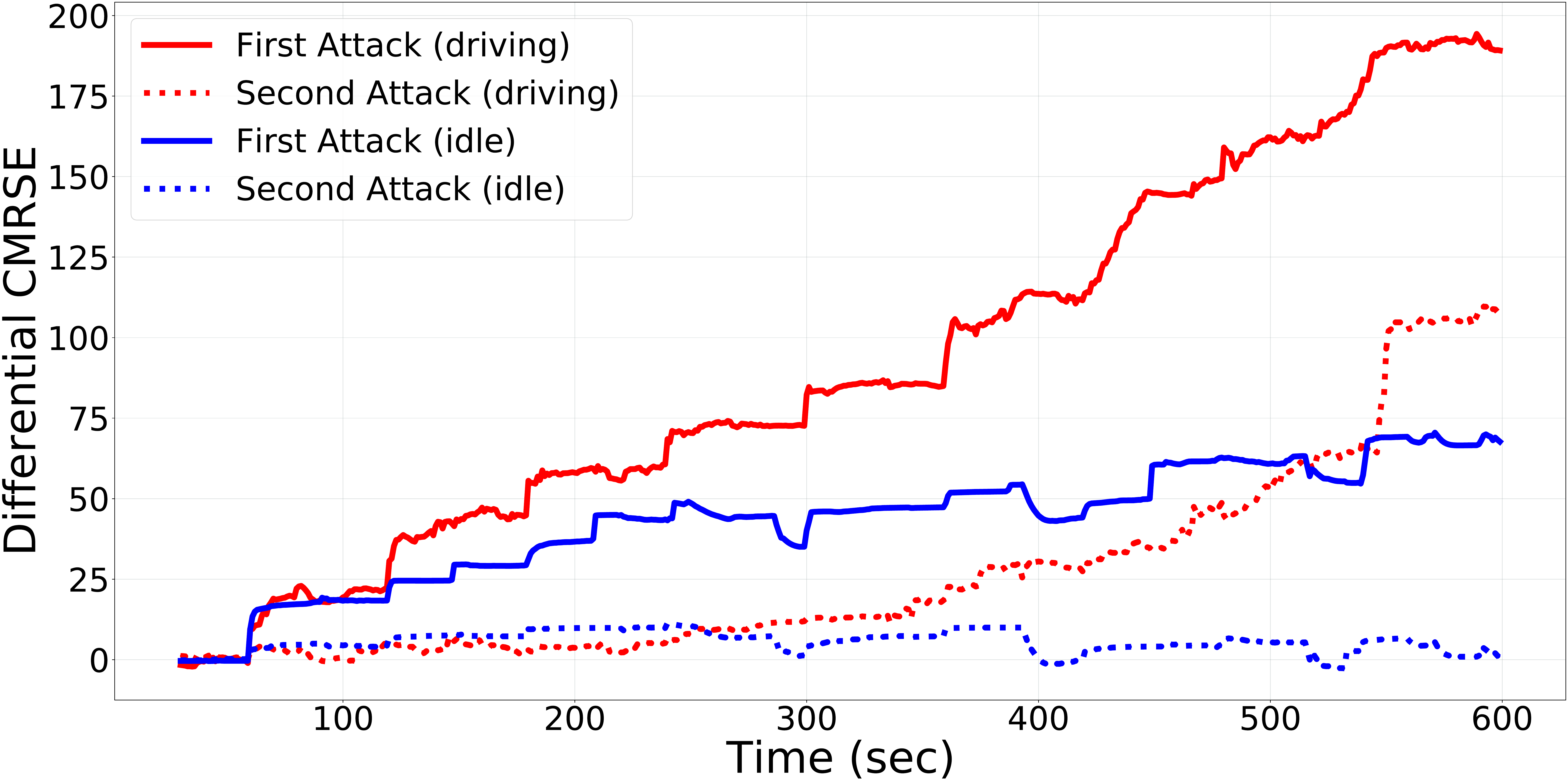}
    \caption{\csirish\ : differential CRMSE of myopic w.r.t. regular behaviors.}
    \label{fig:results_irish}
\end{figure}

\textbf{Results.}
The \textit{LSTM} accumulate errors over time, which can be measured via Cumulative Root Mean Squared Error (CRMSE). We show in Fig.~\ref{fig:results_irish} the difference of such CRMSE for the two considered attack scenarios with respect to clean data (denoted as \emph{differential CMRSE}), during the 10 minutes of the \textit{LSTM} operation. 
We use a full (dotted) line to denote the first (second) attack scenario, whereas red (blue) lines refer to the `driving' (`static') mobility pattern.

We can see that the first attack (full lines) induces a significantly higher differential CRMSE. In comparison, the second attack (dotted lines) is less effective. We find it interesting that, in both scenarios, the malicious CQI occur only 10 times---representing $\smallsim$1\% of the overall sampled data. Even such a small fraction of malicious samples has a major effect, because each sample influences many future predictions. We report in Appendix~\ref{app:irish} the complete time-series of this CS.

Protecting similar ML systems against such myopic attacks is not trivial. The target is an \textit{LSTM} regressor, and its online nature make it difficult to identify adversarial ML countermeasures that do not impair baseline performance. 

%% file: sections/CS/rml.tex
\subsection{\csrml\ : Automatic Modulation Recognition}
\label{ssec:rml}
\textbf{Highlights.} The impact of myopic attacks on \textit{deep} and \textit{shallow} learning is assessed. We also consider \textit{white-/black-box} attacks targeting exactly the same ML system~\cite{usama2021examining, kim2020over}.

\textbf{Target 5G system.}
The ML component focuses on AMR, a task extensively studied by research in 5G (cf. §\ref{ssec:ML5G}). Hence, the data-stream assumes gNB that capture the radio-signals belonging to a given UE, and measure the In-Phase and Quadrature components of such signals. Afterwards, such measurements are sent to a dedicated ML model which must infer the modulation of the corresponding signal. The data used to develop the ML model belongs to the 5G NI tenants, and can be generated either via controlled simulations, or by direct acquisition of physical radio-signals from real UEs.

\textbf{Dataset and Baselines.}
We rely on the \dataset{RML2016.10a} dataset~\cite{o2016radio}.
This well-known dataset includes 220K signals collected for signals at 10 different Signal-to-Noise Ratio (SNR), denoting 11 modulation schemes (i.e., the labels). It is created via the GNU-Radio toolkit, which is appreciated in 5G simulations~\cite{bonati2020open}. The dataset was released in 2016, and \textit{to make it compliant with 5G} we consider only digital modulations. 
Each signals is described by a vector of $128$ pairs of In-Phase/Quadrature (I/Q) measurements. 

As it was done in previous work (e.g.~\cite{flowers2019evaluating, usama2021examining}), we do not apply any preprocessing to the data, hence the features correspond exactly to the raw physical measurements provided in \dataset{RML2016.10a}, thus $F$=($256$ I/Q measurements). To compare shallow with deep learning, we train two multi-class baselines: one is a `shallow' \textit{RF}, the other is exactly the same Deep Neural Network (\textit{DN}) as in~\cite{kim2020over}. We compute accuracy of both baselines: $\text{\textit{Acc}}_{\text{\textit{RF}}}$=$0.82$, $\text{\textit{Acc}}_{\text{\textit{DN}}}$=$0.72$.

\textbf{Attacks.} 
A \textit{myopic} attacker can easily expect that ML systems for AMR analyze I/Q measurements, but cannot reasonably know the exact composition of $F$, hence $\mathcal{F}$=$\overbar{\mathcal{F}}$=$\overbar{F}$=(\text{some I/Q measurements}). Such attacker can artificially generate some noise~\cite{flowers2019evaluating} and affect ML at inference stage.
However, the I/Q measurements are computed at the receiving end (e.g., gNB), which is not accessible. The myopic attacker is hence limited to random and imprecise perturbations, which we simulate by considering four different scenarios. Three involve RsP of randomly chosen measurements: either 25, 50 or 100 (10, 20 or 40\% of $F$, respectively); whereas in the fourth---a worst-case---the RsP affect exactly the 25 most significant measurements for classification. Hence: $\overbar{\mathcal{F}}_{1\text{-}3}$=(25/50/100 rnd I/Q) and $\overbar{\mathcal{F}}_4$=(25 top I/Q).
For each scenario, we craft RsP at 7 increasing intensity levels and attack the baselines. Due to the randomness of $\overbar{\mathcal{F}}_{1\text{-}3}$, we repeat them 20 times. Additional details are in Appendix~\ref{app:rml}.

Let us describe the adversaries considered by related work, all targeting a DN using \dataset{RML2016.10a}.
In~\cite{usama2021examining} a \textit{white-box} attacker has complete knowledge and can freely apply any perturbation to the input data. The setting in~\cite{kim2020over} is more constrained: here, both a \textit{white-box} and a \textit{black-box} attackers want to affect a specific victim (e.g., the owner of a different UE). Hence, they can only apply a perturbation synchronized with the victim's communication channel; such channel can be approximated by sensing the spectrum. All these attackers (in both~\cite{kim2020over} and~\cite{usama2021examining}) know the entire feature set, and their perturbations affect \textit{all} measurements of each signal; by using our notation, for such attackers $\mathcal{F}$=$\mathcal{\overbar{F}}$=$\overbar{F}$=$F$. The attacks in~\cite{usama2021examining} apply a fixed perturbation, whereas those in~\cite{kim2020over} also consider perturbations at 7 increasing intensities.

\begin{figure}[t!]
    \centering
    \begin{subfigure}[t]{0.24\textwidth}
        \centering
        \includegraphics[width=\columnwidth]{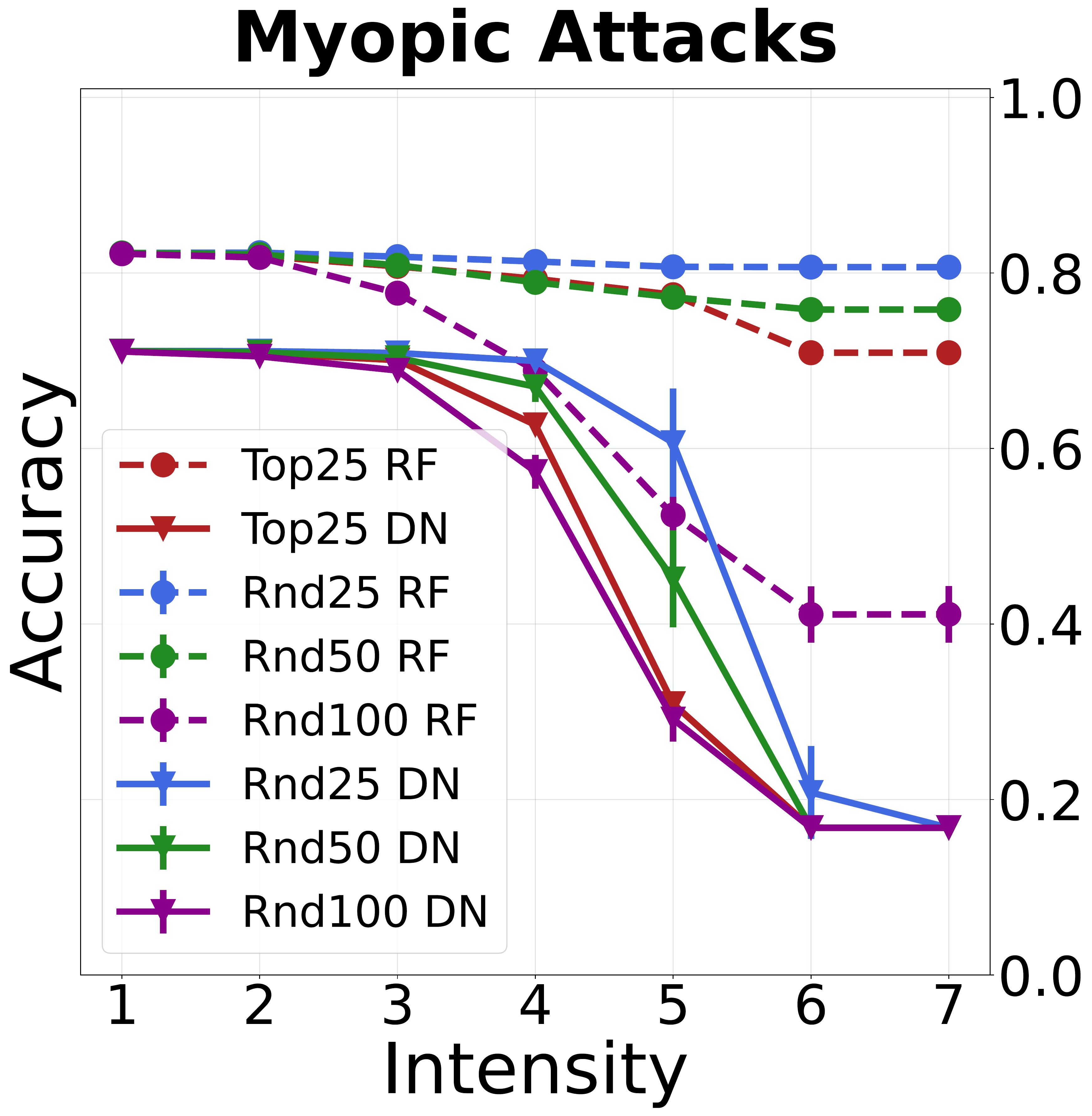}
        \caption{\footnotesize{Deep vs Shallow learning.}}
         \label{sfig:rml_myopic}
    \end{subfigure}%
    ~ 
    \begin{subfigure}[t]{0.24\textwidth}
        \centering
        \includegraphics[width=\columnwidth]{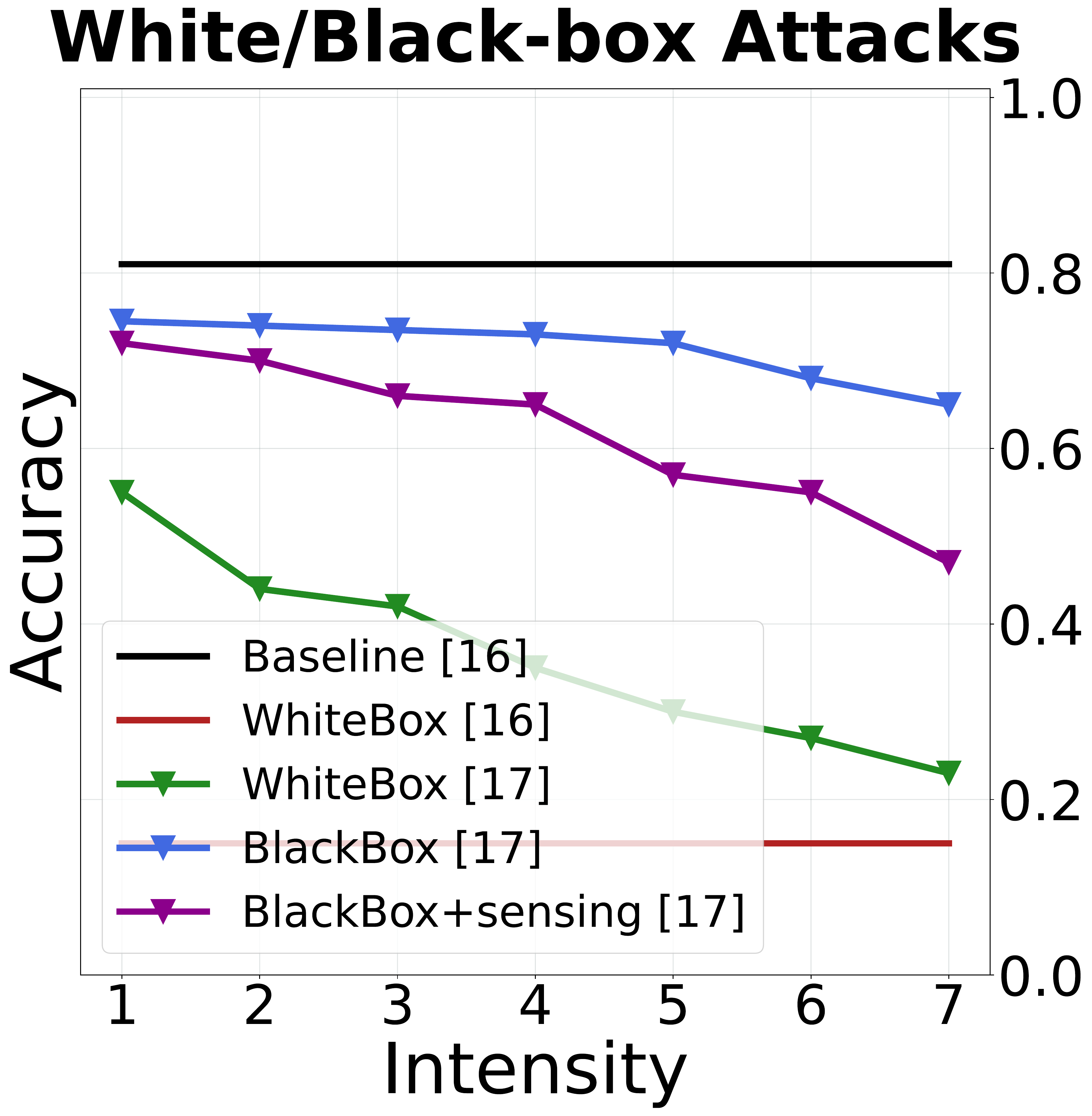}
        \caption{\footnotesize{Attacks in~\cite{kim2020over, usama2021examining}.}}
         \label{sfig:rml_related}
    \end{subfigure}
    \caption{\csrml\ : Myopic Attacks vs White/Black-box Attacks.}
    \label{fig:results_rml}
\end{figure}

\textbf{Results.} We analyse the accuracy degradation attained by myopic attacks against our \textit{DN} and \textit{RF} (Fig.~\ref{sfig:rml_myopic}, the standard deviation of our 20 trials is denoted as a vertical bar on each marker), and compare it with that of attacks in the related work (Fig.~\ref{sfig:rml_related}). These results focus on signals with SNR=10db, which is common in the respective literature.

Shallow learning appears to be more robust than deep learning: in Fig.~\ref{sfig:rml_myopic} the \textit{DN} (full lines) is more affected by the myopic attacks than the \textit{RF} (dotted lines), although both baselines are defeated with random RsP at high intensity. 
Prior work only targeted deep learning, so a fair comparison against such work must focus on the \textit{DN} results in Fig.~\ref{sfig:rml_myopic}. We can see that the attack in~\cite{usama2021examining} (red line in Fig.~\ref{sfig:rml_related}) is devastating but it also assumes an extremely powerful attacker. Conversely, the constrained attacker in~\cite{kim2020over} is much less successful in the black-box setting (blue line in Fig.~\ref{sfig:rml_related}); she becomes more successful if she can obtain more information about the victim, either by sensing the spectrum (purple line), or in the white-box setting (green line).

%% file: sections/CS/pamimo.tex
\subsection{\cspamimo\ : Power Allocation in massive MIMO}
\label{ssec:pamimo}
\textbf{Highlights.} We assess myopic examples in a \textit{single}- and \textit{multi-agent} adversarial setting; the attacks target a \emph{physical} quality metric in 5G networks, the spectral efficiency (SE). We also compare our attacks with a recent work~\cite{manoj2021adversarial}. Finally, an intriguing property of this CS is that the attacker's goal is \emph{reached} through \emph{unsuccessful} adversarial examples.

\textbf{Target 5G system.}
The ML component must estimate the power allocation in mMIMO 5G networks. According to the state-of-the-art, the allocated power can be computed by ML as a function of the position of multiple UE with respect to their serving gNB~\cite{sanguinetti2018deep}. Hence, we consider a system in which the data-stream envisions UEs that, after obtaining their geographical location~\cite{yudnikov2020doppler}, communicates such information to the gNB. The 5G NI uses such information to determine the actual \textit{distance}\footnote{Such distance can be measured by hardcoding the geographical position of the gNB, or by using a dedicated localization service~\cite{schungel2021heterogeneous}.} of all the UEs with respect to their serving gNB. All such distances are then provided to a dedicated ML component deployed within the 5G NI, which estimates how much power should be allocated to each gNB in order to support the attached UEs. Such estimates are finally used by the 5G NI for proper resource management. The excellent results of~\cite{sanguinetti2018deep} show that proficient ML models for power allocation can be trained through entirely simulated data.

\textbf{Dataset and Baseline.}
We use the recent \dataset{PA-mMIMO} dataset~\cite{sanguinetti2018deep} generated by a simulation of 20 UE served by 4 gNBs. Its 335K raw-data samples describe the positions of each UE (x/y coordinate pairs) and the corresponding actual allocated power, making it suitable for 5G experiments~\cite{manoj2021adversarial, sanguinetti2018deep}. 
We replicate the state-of-the-art technique proposed in~\cite{sanguinetti2018deep}: UE report their locations (acquired via GPS) to the 5G NI which uses deep learning to allocate the power of 4 gNBs to 20 UE. We use the source code provided by~\cite{sanguinetti2018deep} to devise our Deep Network (\textit{DN}) baseline whose feature set is $F$=(20 \textit{x/y} pairs). The quality of each predicted power allocation vector is estimated my means of the physical SE metric.

\begin{figure}[!htbp]
    \centering
    \includegraphics[width=0.95\columnwidth]{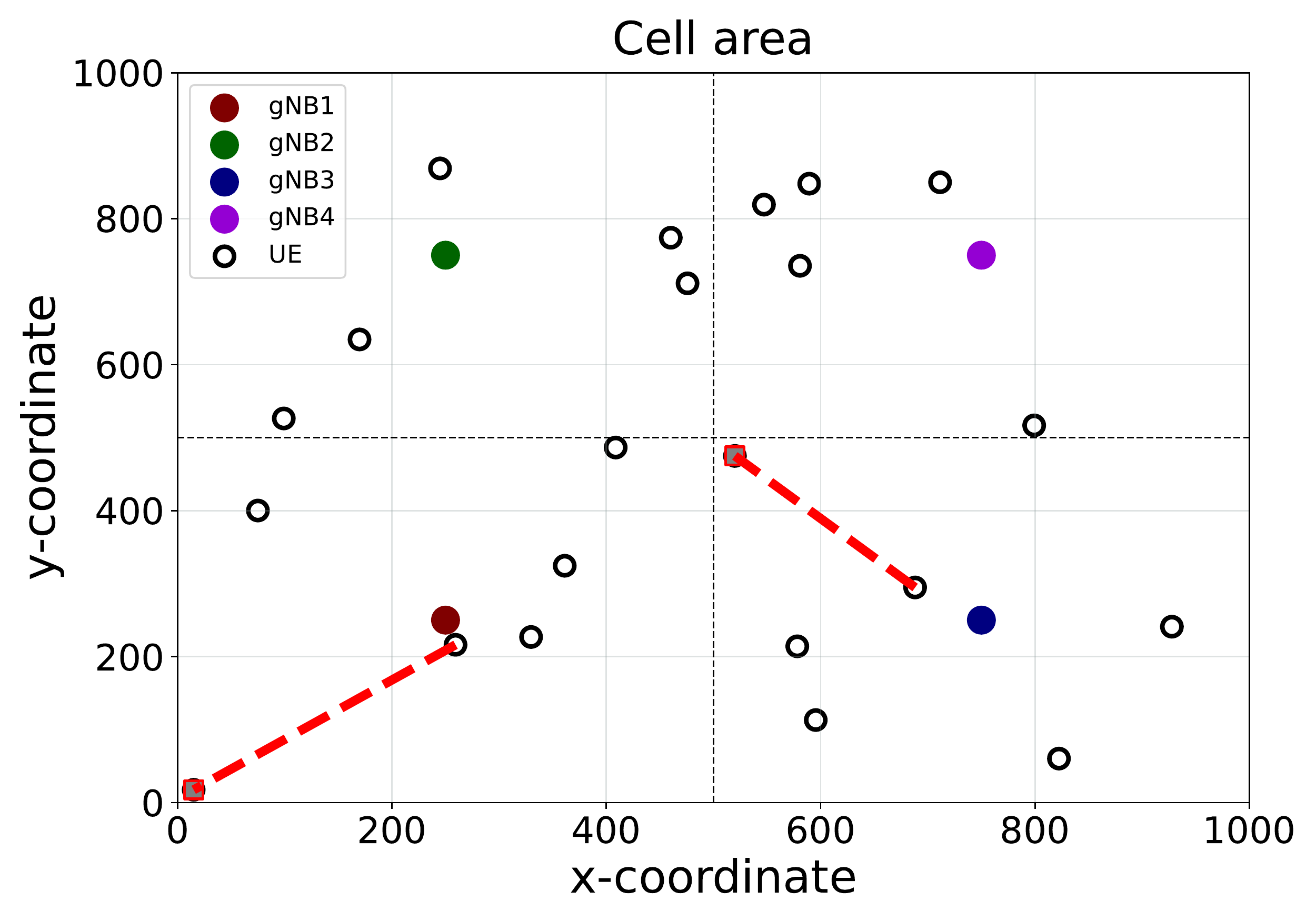}
    \caption{\cspamimo\ : network topology and adversarial movement.}
    \label{fig:pamimo_context}
\end{figure}

\textbf{Attacks.} 
A myopic attacker can expect the usage of location-based information for power allocation. She cannot reasonably know \textit{how many} UEs are served by the system, but she knows that at least her UE is included among them. In this case, the attacker can spoof her geographical position (e.g.,~\cite{zeng2018all,falco2019cybersecurity}), affecting the \textit{DN} at the inference stage; hence, $\overbar{F}$=$\overbar{\mathcal{F}}$=$\mathcal{F}$=(1 \textit{x/y} pair)
An attacker may also collude with a partner in a multi-agent setting by synchronizing their attacks, increasing their impact on the whole system; hence, $\overbar{F}$=$\overbar{\mathcal{F}}$=$\mathcal{F}$=(2 \textit{x/y} pairs). Such scenarios are depicted in Fig.~\ref{fig:pamimo_context}, showing an mMIMO network with 4 cells and 4 gNBs (filled dots) each serving 5 UEs (empty dots), for a total of 20 UEs.
For the single-agent setting, we consider an attacker whose UE is served by gNB1; for the multi-agent setting, we consider an additional attacker whose UE is served by gNB3. Myopic attackers can fake their location, but they cannot determine the exact values that maximizes their impact. They may, however, expect that greater (faked) distances from the gNB cause more power to be allocated to their UE. We simulate such behavior by ``moving'' the attackers' UE away from the gNB. This is done via RsP that alter the \textit{x/y} coordinates of the myopic UE(s), so that the distance from the serving gNB increases in 8 steps in a range of [0-300] within the same cell (shown with a dotted trajectory in Fig.~\ref{fig:pamimo_context}).

\begin{figure}[!htbp]
    \centering
    \includegraphics[width=0.95\columnwidth]{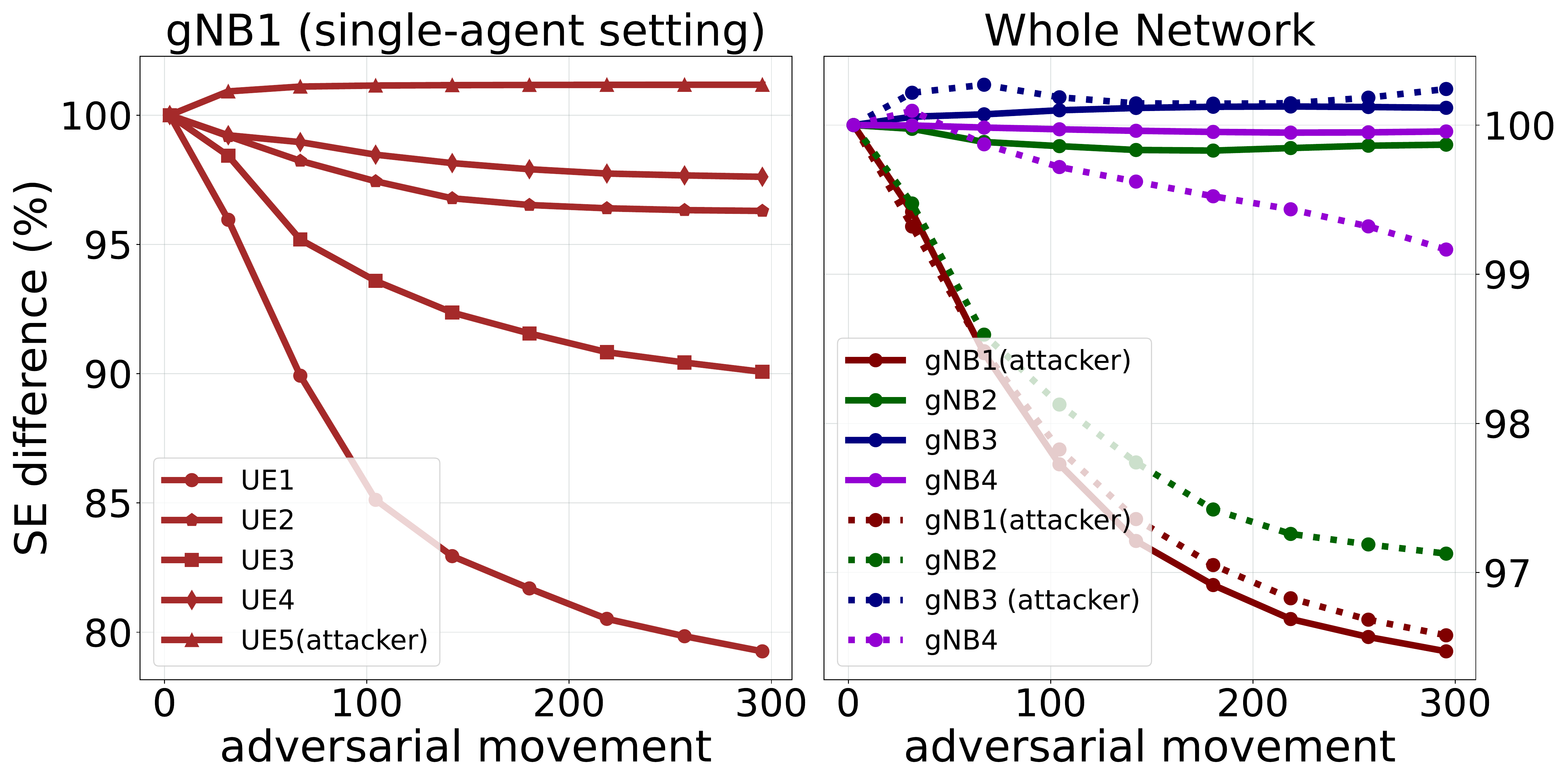}
    \caption{\cspamimo\ : attack impact on a single cell and whole network, measured as the difference (in \%) of the SE as a function of the adversarial movement. 
    The left plot shows the effects of the single-agent setting on the specific cell served by the closest gNB1; the right plot shows the effects of both the single- and multi-agent setting (full and dotted lines, respectively) on the whole system, where the values are averaged among all UEs in each cell.}
    \label{fig:pamimo_attack}
\end{figure}

\textbf{Results.} 
We show the impact of the attack on the SE in Fig.~\ref{fig:pamimo_attack}. Because the attacking UEs never truly move the correct power allocation should not change and the corresponding SE should remain the same (i.e., at 100\%).
In the single-agent setting (solid lines), we can see from the left plot that the SE of \textit{all} UEs in the gNB1 cell is affected: the attacker (UE5) gains about 3\% SE at the expense of decreasing the SE of other UEs by 3\%--20\%. As a consequence, the 5G NI allocates more power to serve the attacker's UE5 and less for the other UEs, preventing optimal QoS. This setting also slightly affects the UEs in the neighboring cells, as shown in the right plot (with green, blue, purple solid lines).
In the multi-agent setting (dotted lines), the impact of an attack is more complex. The \emph{average} SE of a cell, shown in the right plot in dotted lines, can decrease as well as slightly increase as a consequence of an attack. 
What happens is that the attackers' UEs (in gNB1 and gNB3) gain more SE than the other UEs in their cells. This leads to an overall degradation of SE in the whole network (confirmed by gNB2 and gNB4 which have only benign UEs). 

This CS showcases another intriguing property of the myopic threat model: the attacker can damage the 5G NI even if all adversarial examples are unsuccessful, from the ML point of view. Indeed, the generated perturbations elicit the \emph{correct} response from the \textit{DN} (as opposed to the wrong response expected from adversarial examples). However, the system QoS is still damaged because the power allocated to serve each UE in the network is incorrect. This happens because the \textit{DN} approximates power allocation as a multidimensional function depending on the positions of \textit{all} UEs in the network. Due to the novelty of this finding it is not immediately clear how such attacks could be prevented at the ML level. 

We now compare our myopic attacks against a very recent work that targets exactly the same \textit{DN} with white/black-box attacks. In~\cite{manoj2021adversarial}, the attacker is very powerful: she controls and can freely ``move'' \textit{all} 20 UEs of the system. Furthermore, the capability to query the \textit{DN} (black-box) by manipulating the entire environment, or the full knowledge (white-box) can be exploited to find the optimal attack vector causing an incorrect prediction. Specifically, by moving some UEs \textit{within centimeters}, the SE can be decreased up to 60\% and 20\% in white- and black-box settings, respectively. We found it surprising that our 'primitive' myopic attacker controlling one UEs could cause a similar decrease in SE (e.g., UE3 in the left plot of Fig.~\ref{fig:pamimo_attack}) as a black-box attacker having access to the 5G NI and controlling all 20 UEs.

%% file: sections/CS/deepslice.tex
\subsection{\csdeepslice\ : A secure by design ML system against myopic attacks}
\label{ssec:deepslice}
\textbf{Highlights.} We showcase a state-of-the-art ML system that \textit{cannot} be affected by a myopic attacker.

\textbf{Target 5G system.}
The ML component focuses on network slicing but---differently from \csctu\ ~(§\ref{ssec:ctu})---the objective is to determine the slice assignment by using KPIs \textit{contained in the 5G NI}~\cite{thantharate2019deepslice}. Therefore, this CS entails a data-stream in which UEs, served by a given gNB, leverage the connectivity of the 5G NI. However, the decision to allocate more (or less) resources for such gNB (and corresponding UEs) depends on KPI related to the current state of the 5G NI (e.g., the time of the day, or the packet delay budget). Such slicing policies can be implemented by training a deep learning classifier on data provided by the 5G NI tenants, specifying which `slice' should be chosen according to the status of the 5G NI~\cite{thantharate2019deepslice}.

\textbf{Dataset and Baseline.}
We use the \dataset{DeepSlice} dataset~\cite{thantharate2019deepslice}, released in 2019 by following and characteristics of the three 5G use cases: eMBB, MMTC, URLLC (which are the three possible `slices'). It was also used in~\cite{thantharate2020secure5g} to propose a defensive mechanism against DoS attacks. 
We replicate the same multi-class \textit{DN} as in~\cite{thantharate2019deepslice} using the same feature set $F$ (reported in Table~\ref{tab:deepslice_features} in Appendix~\ref{app:deepslice}). Our \textit{DN} matches the perfect Accuracy of~\cite{thantharate2019deepslice}. 

\textbf{Attacks.} 
A myopic attacker that is aware of~\cite{thantharate2019deepslice} may want to target a similar ML system to disrupt resource allocation. Such attacker can infer the entire feature set of~\cite{thantharate2019deepslice}, hence $\mathcal{F}$=$F$. However, the only features in $\mathcal{F}$ that a myopic attacker can influence are those related to time (e.g., \textit{hour} and \textit{day}): the other features (e.g., \textit{PktDelayBudget}) depend on the 5G specifications and can only be modified by the 5G NI tenants, and are not controllable by the myopic attacker. Hence, $\overbar{\mathcal{F}}$=$\overbar{F}$=(\textit{hour}, \textit{day}). We apply the RsP by considering any combination of the \textit{day} and \textit{hour}.

\textbf{Results.}
Not a single myopic example was successful against the baseline \textit{DN}.
This is an example of a ``secure-by-design'' system against myopic attacks\footnote{We stress that the notion of ``secure-by-design'' should be contextualized. In our case, we use such notion only to refer to systems that cannot be affected by myopic attacks (which are, by definition, launched from outside the 5G NI). Of course, the ML system in \csdeepslice\~ can still be attacked via other forms of adversarial attacks---if their requirements are met by the adversary.}, because the features relevant for classification (e.g., \textit{PktDelayBudget}) can be influenced only with direct access to the 5G NI. This can be possible for an `insider', but such assumption would violate our threat model. 
Regardless, for those interested in such `insider myopic attacks', we showcase these attempts in Appendix~\ref{app:deepslice}, which also includes additional details as well as the evaluation of ML-specific countermeasures.

%% file: sections/discussion.tex
\subsection{Discussion}
\label{ssec:discussion}

Let us finalize our evaluation by discussing some crucial remarks from a practical point of view. 

\textbf{Are myopic attacks undefeatable?} The attacks in \csctu{}, \cspamimo{} and \csirish{} are successful and ML countermeasures are difficult to apply or not very effective. However, as shown by \csdeepslice{}, some ML systems are secure-by-design against some forms of myopic attacks. Moreover, \cselasticmon{} clearly shows that some well-known adversarial ML countermeasures can mitigate or completely defuse our myopic attacks. However, to prevent the attacks in \cselasticmon{}, it is necessary to \textit{predict} either the features attacked (for feature removal) or the specific RsP (for adversarial training): doing this requires a proactive approach, which is the one endorsed by our paper. Finally, we hope that our paper will inspire future work aimed at \textit{specifically} countering myopic attacks (with a favorable tradeoff).

\textbf{Are myopic attacks more dangerous than existing attacks against ML?} The white/black-box attacks of~\cite{manoj2021adversarial} and those in~\cite{usama2021examining, kim2020over} (cf. \cspamimo{} and \csrml{}) are more disruptive, but they also require a higher resource investment to be staged; for instance, our myopic attacks only require a rough knowledge of the feature set, which can be obtained by reading technical reports or scientific papers. It is hence crucial that both circumstances are taken into account when testing ML components deployed in critical infrastructures.

\textbf{Can myopic attacks target different network infrastructures?} Yes, but---to be viable---only if such infrastructures (i)~use ML, (ii)~are open, and (iii)~are subject to granular SLA (cf. §\ref{ssec:viability}). If a network is not open, then an attacker cannot easily leverage multiple UEs to stage myopic attacks; without granular SLA, the damage potential of myopic attacks is decreased; and without ML it is simply impossible to conceive attacks based on adversarial examples.

\textbf{Do the CS represent feasible adversarial attacks?} Yes. Our CS are based on the predictions of ML components to adversarial examples, and are adversarial attacks by definition (cf. Eq.~\ref{eq:AA}). Such examples are created via RsP manipulations that are possible by attackers who physically own their UE(s). To provide a broad assessment our RsP have varying intensity, because real attackers are not interested in crafting the `minimal' perturbation. RsP with low intensity may have little effect, but higher intensities can be disruptive (e.g., Fig.~\ref{fig:pamimo_attack}).

\textbf{Can myopic attacks be blocked via non-ML protection mechanisms?} Yes. For instance, the attacks in \cspamimo{} would be defused by ensuring the correctness of the position reported by UE. However, such mechanisms are not cheap to implement\footnote{They may require stateful analyses of diverse data-sources (e.g.,~\cite{jansen2018crowd}).}, can be broken (e.g.,~\cite{lakshmanan2021stealthy}), and may induce overheads that would nullify the advantages provided by ML in the 5G NI.

\textbf{Is the real 5G NI endangered by myopic attacks?} Yes. In our CS, we attack ML prototypes based on state-of-the-art techniques for 5G networking. The deployed 5G NI will behave differently, since its ML components are trained on different (private) datasets, and the processing pipelines may include additional (proprietary) mechanisms---all of which are at the discretion of the 5G NI tenants and/or under NDA and hence not available for research. However, we interviewed several 5G telcos who acknowledged that our CS are indeed feasible and represent a threat that must be taken into account.

\textbf{Are our findings generalizable?} We acknowledge that, with the exception of \csrml{}, we only attack a single ML model per case study---hence, we do not claim that ``every ML model deployed in practice is vulnerable to myopic attacks''. However, all our CS revolve around ML methods proposed by the state-of-the-art, and hence represent those that are more likely to be deployed in the real 5G NI (because they provide the highest performance). Nonetheless, we hope that our paper will inspire future works that will consider different ML models: perhaps, some ML techniques will provide a slightly lower baseline performance, but are naturally robust against myopic perturbations. We believe that such an outcome would be \textit{valuable} for real ML deployments.

%% file: sections/6-related.tex
\section{Related Work}
\label{sec:related}

Limited attention has been given to adversarial examples in 5G networking. 
Some papers provide a broad overview of the security risks of ML-powered wireless communications~\cite{sagduyu2020wireless, kafle2018consideration, ahmad2020challenges}; others propose ML-based security mechanisms in 5G (e.g.,~\cite{wang2020pilot, thantharate2020secure5g}). An exhaustive survey is~\cite{suomalainen2020machine}. 
None of these works evaluate or propose original threat models. 

Some papers have little in common with 5G networking. The authors of~\cite{usama2018adversarial, Ibitoye:Analyzing, monge2019traffic} consider attacks against cyber detectors.
Similarly,~\cite{wu2020defense, li2019desvig,pajola2019threat,qiu2020artificial} target ML systems for computer vision and autonomous driving. All these tasks---despite being strongly linked with 5G~\cite{arjoune2020artificial}---are unrelated to networking functions. Finally, some orthogonal studies consider the usage of ML as an \textit{offensive} mechanism~\cite{luo2020attackers, sagduyu2019adversarial, erpek2018deep}.

Let us directly compare our paper with closely related work, summarized in Table~\ref{tab:related}. For each paper, we report the threat model (\whitecirc\ and \blackcirc\ denote white- and black-box attacks); whether the perturbations adhere to some constraints; the assessment of defenses; and the usage of public data.

\input{sections/table_related}

Some papers consider white-box attackers with complete knowledge of the target ML model that can freely apply any perturbation (without providing any justification)~\cite{usama2021examining, manoj2021adversarial, usama2019adversarial}. Other works consider more realistic scenarios where the perturbations are subject to physical constraints (e.g.~\cite{flowers2019evaluating, kim2020over, kim2020adversarial, restuccia2020generalized, hameed2020best}), despite considering attackers with full knowledge of the ML system or that can observe its output.

To be viable, all feedback-based strategies pose an additional requirement: the attacker must be certain that the feedback corresponds exactly to the ML output. By using the notation in §\ref{ssec:myopic}, these attackers must be aware that $N(M\!+\!I)$=$M(F_x)$, and that such $N(M\!+\!I)$ corresponds to the feedback of the 5G NI. Obtaining such certainty is tough without access to the infrastructure hosting the target ML systems. Another possibility is if the attacker owned \textit{all} the UEs served by the 5G NI, allowing to monitor the results of all their interactions--which is clearly unfeasible.

The authors of~\cite{davaslioglu2019trojan} apply specific manipulations to the training set (i.e., trojaning); having such direct access to training data is unlikely in critical environments. More viable poisoning attempts are found in~\cite{sagduyu2019iot, shi2018spectrum} where the attacker has less control and must first `sense' the spectrum. A similar and viable strategy is the jamming attack in~\cite{sadeghi2019physical}: the attacker perpetually inspects the communication channel to replicate (and then disrupt) a ML model. This procedure can only work against ML systems trained on the exact data distribution captured during the sensing activity of the attacker. 

All the adversarial attacks against the 5G NI considered by past work are agreeably not impossible; however, launching them \textit{in real 5G contexts} requires information obtainable only through direct access to the 5G NI. This can happen either as a result of an insider threat~\cite{joshi2020insider}, or via a prolonged and resource intensive APT campaign~\cite{Brewer:Advanced}. We do not claim that our threat model is the only way to realistically attack the 5G NI via adversarial examples.

With respect to existing work, we point out the exposure of the 5G NI to a more affordable but still dangerous adversarial attack strategy, which we formalize with our novel threat model. Moreover, no previous paper has addressed the importance of conducting realistic and generic assessments of adversarial attacks against 5G NI, and only few consider countermeasures. In addition, previous papers usually focus on a single dataset (most notably, \dataset{RML2016.10a}) which sometimes is not publicly released, preventing reproducibility; in contrast, we consider 6 case studies all based on open data, representing the largest assessment of adversarial attacks against the 5G NI. Finally, our original evaluation framework will hopefully facilitate the assessment of adversarial attacks against 5G ML systems---at least until such systems become available for research purposes.

%% file: sections/table_related.tex
\begin{table}[!htbp]
    \centering
    \caption{Prior works on adversarial ML in 5G networking.}
    \label{tab:related}
    \resizebox{0.9\columnwidth}{!}{
        \begin{tabular}{c|c?c|c|c|c|}

             \rotatebox{0}{\textbf{Paper}} & \rotatebox{0}{\textbf{Year}} & 
             \rotatebox{0}{ \begin{tabular}{c} \textbf{Attacker} \end{tabular}} &
             \rotatebox{0}{ \begin{tabular}{c} \textbf{Constr.} \\ \textbf{Perturb.} \end{tabular}} &
             \rotatebox{0}{ \begin{tabular}{c} \textbf{Defenses} \end{tabular}} &
             \rotatebox{0}{ \begin{tabular}{c} \textbf{Public} \\ \textbf{Data} \end{tabular}} \\
            \toprule
            \cite{sadeghi2018adversarial} & 2018 & \whitecirc~/ \blackcirc & \cmark & \xmark & 1 \\ \hline
            \cite{shi2018spectrum} & 2018 & \blackcirc & \cmark & \xmark & \xmark \\ \hline
            \cite{sadeghi2019physical} & 2019 &  \whitecirc~/ \blackcirc & \cmark & \xmark & \xmark \\ \hline
            \cite{kokalj2019targeted} & 2019  & \whitecirc~/ \blackcirc & \cmark & \cmark & 2 \\ \hline
            \cite{bair2019limitations} & 2019  & \whitecirc & \cmark & \xmark & 1 \\ \hline
            \cite{kokalj2019adversarial} & 2019  & \whitecirc & \cmark & \cmark  & 1 \\ \hline
            \cite{sagduyu2019iot} & 2019 & \blackcirc & \cmark & \cmark  & \xmark \\ \hline

            \cite{davaslioglu2019trojan} & 2019 & \blackcirc & \cmark & \cmark & \xmark \\ \hline
            
            \cite{sagduyu2019adversarial} & 2019 & \blackcirc & \cmark & \cmark & \xmark \\ \hline
            
            \cite{usama2019adversarial} & 2019 & \whitecirc & \xmark & \cmark & \xmark \\ \hline
            
            \cite{flowers2019evaluating} & 2019 & \whitecirc & \cmark & \xmark & 1 \\ \hline
            
            \cite{restuccia2020generalized} & 2020 & \whitecirc~/ \blackcirc & \cmark & \xmark & 1 \\ \hline

            \cite{kim2020adversarial} & 2020 & \whitecirc~/ \blackcirc & \cmark & \xmark & 1 \\ \hline
            
            \cite{hameed2020best} & 2020 & \whitecirc~/ \blackcirc & \cmark & \cmark & 1 \\ \hline

            \cite{usama2021examining} & 2021 & \whitecirc~/ \blackcirc & \xmark & \xmark & 2 \\ \hline
            
            \cite{kim2021adversarial} & 2021 & \whitecirc~/ \blackcirc & \xmark & \xmark & \xmark \\ \hline
            
            \cite{manoj2021adversarial} & 2021 & \whitecirc~/ \blackcirc & \xmark & \xmark & 1 \\ 
            
            \midrule
            \multicolumn{2}{c?}{Ours} & myopic & \cmark & \cmark & 6 \\ 
            \bottomrule
        \end{tabular}
    }
\end{table}

%% file: sections/7-conclusions.tex
\section{Conclusions}
\label{sec:conclusions}

The security of the 5G Network Infrastructure (NI) is of paramount importance due to its critical role in the current and future society. Empowering such infrastructure with ML exposes it to the risk of adversarial examples, which have not received adequate treatment in this context. The 5G paradigm enables a new class of harmful adversarial ML attacks with a low entry barrier, which \textit{cannot} be formalized with existing adversarial ML threat models. Furthermore, such vulnerabilities must be \textit{proactively} assessed, but the early stage of ML in SA 5G makes such evaluations challenging for research.

In this paper, we propose the first threat model that is specific to adversarial ML attacks against the 5G NI. Our `myopic' threat model describes viable attacks that can inflict damage (physical as well as monetary) to the 5G NI tenants. Moreover, we provide an original security evaluation framework based on open source data, enabling realistic assessment of adversarial examples against state-of-the-art ML systems. Both our threat model as well as our framework are \textit{agnostic} of the specific function solved by ML in the 5G NI, and hence cover even yet to be conceived applications of ML in SA 5G.

We apply our framework to evaluate the proposed myopic threat model, and analyze six case studies where we target state-of-the-art ML systems for 5G NI. We show that 5 out of 6 systems can be broken with our myopic attacks which can influence both the training and inference stages; the attacks can also simultaneously affect single devices and the entire environment and can be amplified by multi-agent strategies or acquisition of more UE. All of these circumstances can cause damage to the 5G NI tenants due to SLA violations. Our attacks may have a smaller success rate than prior black-/white-box attacks but do not require any compromise of the 5G NI. Finally, we showcase a ML system which is immune to our attacks---by design.

This paper can inspire many research directions, such as case studies using the real ML elements in SA 5G---when they become available for adversarial ML research purposes.
The proposed framework also allows the discovery of existing datasets that are usable for realistic adversarial ML evaluations.
Another possibility is estimating the monetary damage of myopic attacks as a consequence of SLA violations.

We have formalized and assessed a new class of attacks against ML systems in SA 5G. Yet, at this point in time, foreseeing the possible incarnations of such systems and evaluating their security risks is difficult---but necessary to ensure their reliability for our society. We hope that our contribution will serve as an important step towards secure ML for 5G networking.

%% file: main.bbl

%% file: biographies/bio.tex
\vspace{-5em}
\begin{IEEEbiography}
  [{\includegraphics[width=1in,height=1.25in,keepaspectratio]{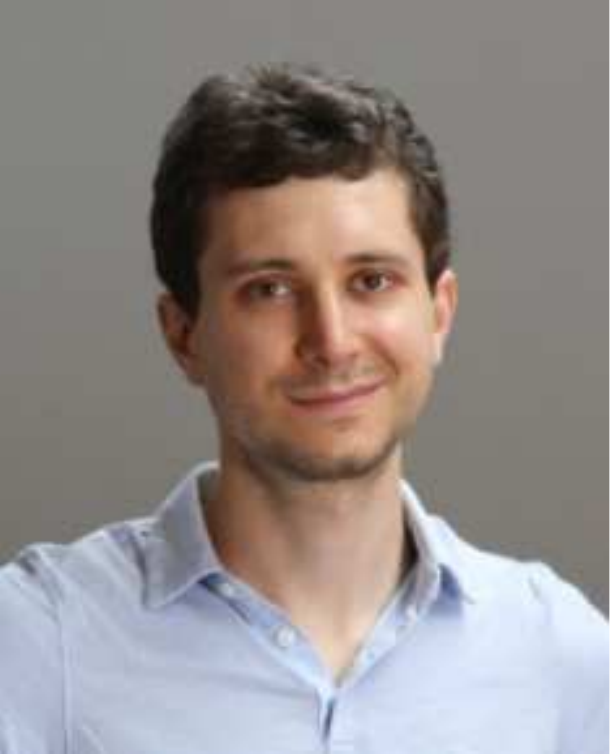}}]{Giovanni Apruzzese} is a Post-Doctoral researcher within the Institute of Information Systems at the University of Liechtenstein since 2020. He received the PhD Degree and the Master's Degree in Computer Engineering (summa cum laude) in 2020 and 2016 respectively at the University of Modena, Italy. In 2019 he spent 6 months as a Visiting Researcher at Dartmouth College (Hanover, NH, USA) under the supervision of Prof. VS Subrahmanian. His research interests involve all aspects of big data security analytics with a focus on machine learning, and his main expertise lies in the analysis of Network Intrusions, Phishing, and Adversarial Attacks.
  
\end{IEEEbiography}
\vspace{-5em}
\begin{IEEEbiography}
 [{\includegraphics[width=1in,height=1.25in,clip,keepaspectratio]{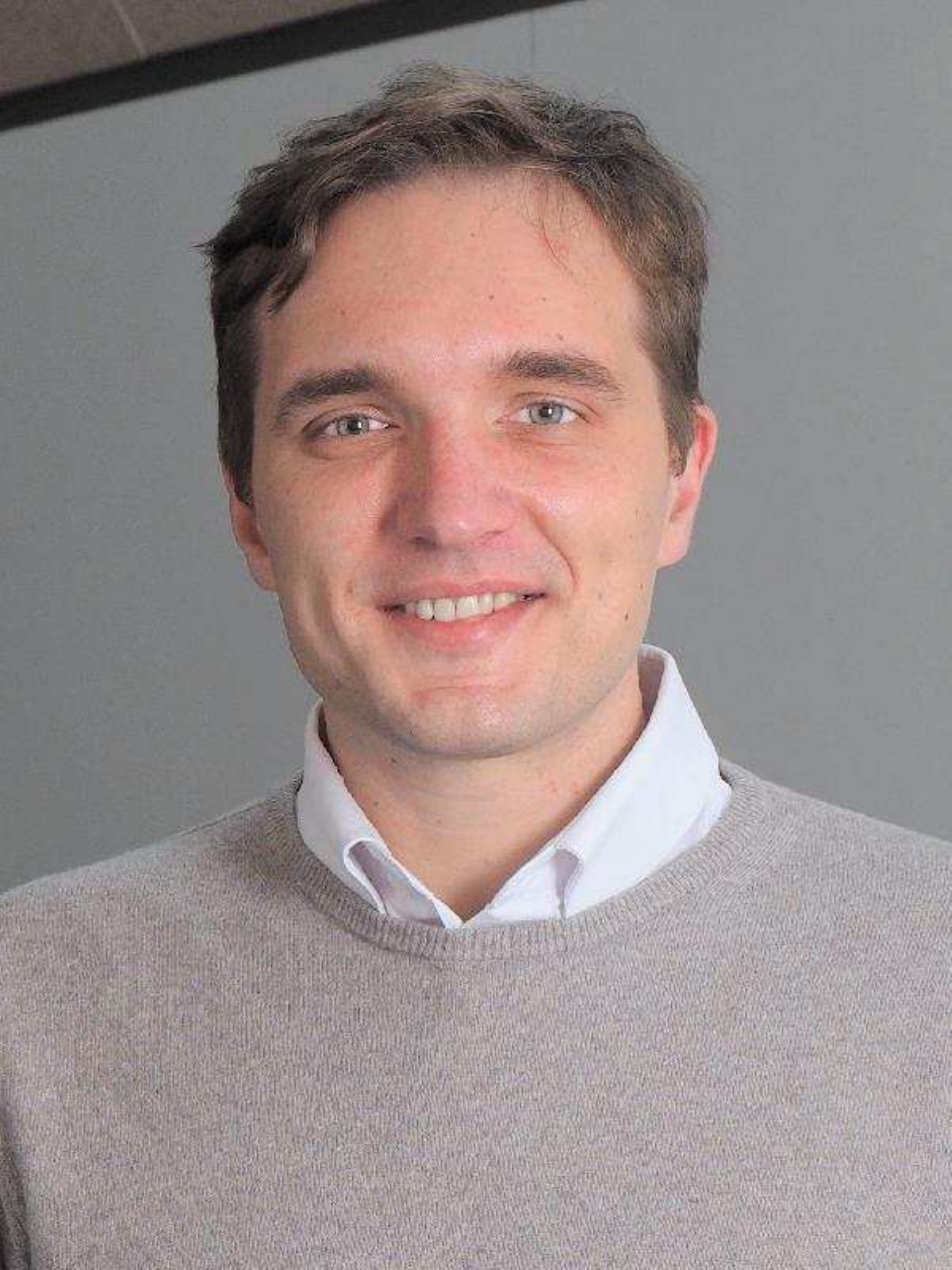}}]{Rodion Vladimirov} is a PhD student within the Institute of Information Systems at the University of Liechtenstein since 2021. He received the BSc. Degree in Mathematical Methods in Economics at Ural State Technical University in Russia (2015) and the MSc. Degree in Finance at the University of Liechtenstein (2020). His main research is focused on the security analysis of AI-systems in 5G networks and includes the investigation of attacks against Machine Learning components in critical infrastructures and the design of appropriate countermeasures.
 
\end{IEEEbiography}
\vspace{-5em}
\begin{IEEEbiography}
 [{\includegraphics[width=1in,height=1.25in,clip,keepaspectratio]{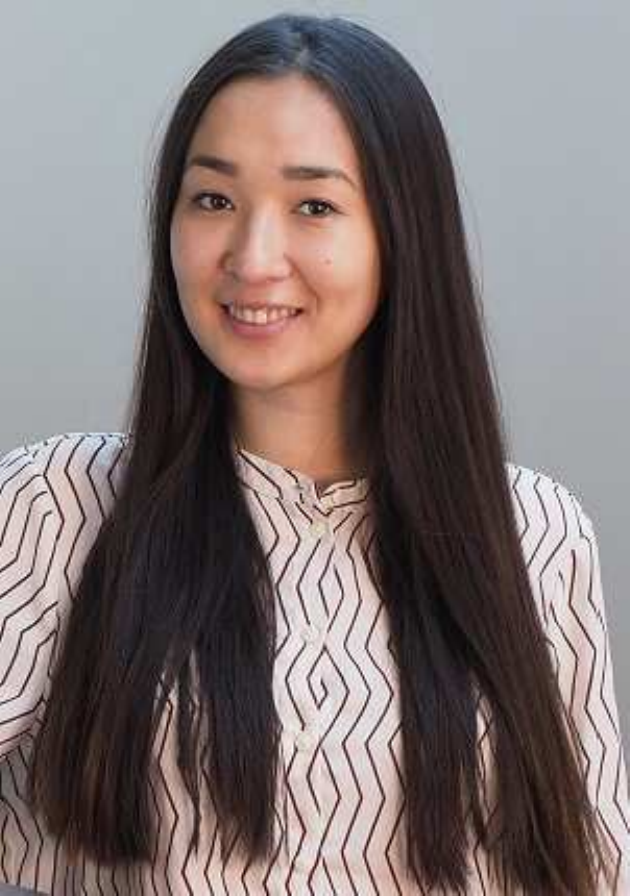}}]{Aliya Tastemirova} received the Master's Degree in Information Systems at the University of Liechtenstein in 2021. Within the same university, she worked as a Student Assistant at the Institute of Information Systems from 2020 to 2021. She is now a Software Developer at Odoo.
\end{IEEEbiography}
\vspace{-8em}
\begin{IEEEbiography}
 [{\includegraphics[width=1in,height=1.25in,clip,keepaspectratio]{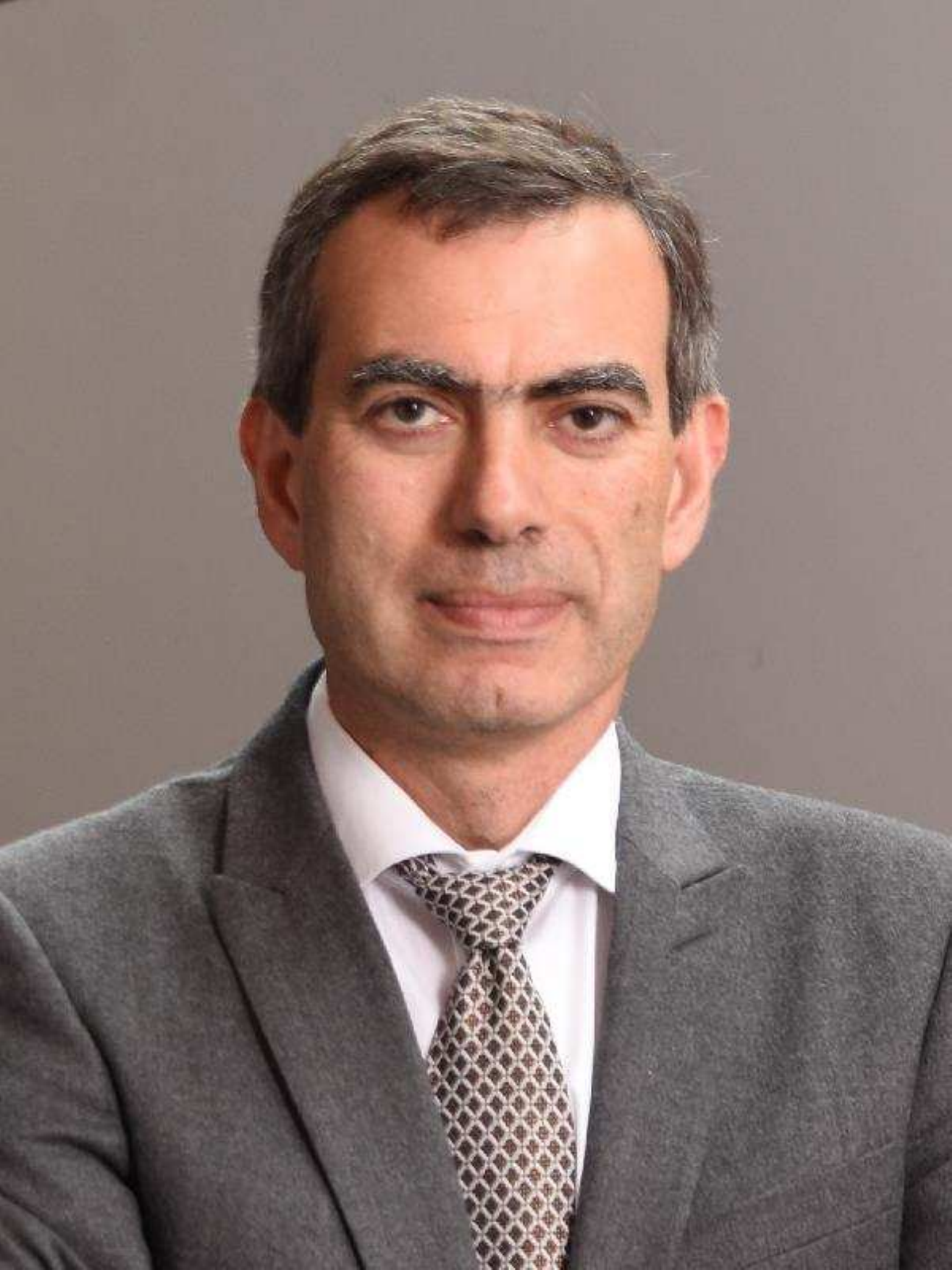}}]{Pavel Laskov} is Full Professor at the University of Liechtenstein and head of the Hilti Chair of Data and Application Security. He received PhD in computer science at the University of Delaware in 2001 and held research and teaching positions at the Fraunhofer Institute FIRST, University of Tuebingen and Huawei European Research Center. His research is focused on the development of techniques for detection and mitigation of security incidents, especially using custom-built AI techniques. As one of the pioneers of research on AI security, Pavel Laskov co-designed the first proof-of-concept attacks against mainstream AI algorithms such as neural networks and Support Vector Machines.  
 
\end{IEEEbiography}

%% file: sections/A1-details.tex
\section{Experimental Testbed}
\label{app:testbed}

\input{sections/table_cs}

All our experiments are performed on a machine equipped with an Intel Xeon W-2195 CPU with 36 cores, 256GB RAM, 2TB SSD NVMe, and Nvidia Titan RTX GPU. The implementation leverages Python3 and popular ML libraries: scikit-learn, Keras, and Tensorflow. 
More details on each CS are found in the remainder of this appendix.

\input{sections/app_CS/app_ctu}

\input{sections/app_CS/app_elasticmon}
\input{sections/app_CS/app_irish}
\input{sections/app_CS/app_rml}

\input{sections/app_CS/app_deepslice}

%% file: sections/table_cs.tex
The datasets chosen for our CS are directly related to the exemplary network functions in 5G infrastructures described in §\ref{ssec:ML5G}. An overview of our CS and corresponding datasets is provided in Table~\ref{tab:datasets}. For each dataset, we report whether it contains real or synthetic data, its size in samples (after preprocessing), how many features are extracted to devise the corresponding ML methods and how many classes are considered (some CS address a regression problem denoted as ``Regr''). Detailed descriptions and motivations for these datasets are provided in each CS.

\begin{table}[!htbp]
    \centering
    \caption{Summary of the datasets of our Case Studies.}
    \label{tab:datasets}
    \resizebox{\columnwidth}{!}{
        \begin{tabular}{c|c|c|c|c|c|c}

            \multirow{2}{*}{\textbf{CS\#}} &
            \multicolumn{5}{c|}{\textbf{Dataset Information}} & \multirow{2}{*}{\begin{tabular}{c} \textbf{5G Network} \\ \textbf{ML Function} \end{tabular}} \\ \cline{2-6}
            & \textbf{Name} & \textbf{Origin} & \textbf{Size} & \textbf{Classes} & \textbf{Features} & \\ 
            \toprule
            
             \csctu\ & \dataset{CTU13}~\cite{Garcia:CTU} & Real & $5.5$M & 2 & 13 & Slicing \\ \midrule

             \cselasticmon\ & \dataset{ElasticMon}~\cite{vasilakos2020elasticsdk} & Synt & $27$K & Regr. & 16 & CQI Pred. \\ \midrule
             
             \csirish\ &  \dataset{Irish 5G}~\cite{raca2020beyond} & Real & $2.4$K & Regr. & 2 & CQI Pred. \\ \midrule

             \csrml\ & \dataset{RML2016}~\cite{o2016radio} & Synt & $120$K & 6 & 256 & AMR \\
            \midrule
            
            \cspamimo\ & \dataset{PA-mMIMO}~\cite{sanguinetti2018deep} & Synt & $335$K & Regr. & 20 & Pow. Alloc. \\
             \midrule
            
            \csdeepslice\  &  \dataset{DeepSlice}~\cite{thantharate2019deepslice} & Synt & $63$K & 3 & 8 & Slicing\\

            \bottomrule
            
        \end{tabular}
    }
\end{table}

%% file: sections/app_CS/app_ctu.tex
\subsection{\csctu\ : dataset, preprocessing, RsP}
\label{app:CTU}
This CS is based on the \dataset{CTU13} dataset~\cite{Garcia:CTU}, collected in a large university campus (almost 300 internal hosts), and provided both as raw packet captures (PCAP) as well as NetFlows generated via Argus\footnote{\url{https://openargus.org/}}. Despite being collected in 2013, the considered network is large, has a high bandwidth, and NetFlows are still widely employed today and can be used for slicing operations in 5G NI (as seen in~\cite{coronado2019flow, li2018deep}).
The \dataset{CTU13} dataset is popular in the intrusion detection community for the development of ML botnet detectors (e.g.,~\cite{Stevanovic:Botnet}), a task unrelated to the scope of this paper. However, we observe that \dataset{CTU13} provides ground truth on 4 distinct classes of communications: \textit{active} (e.g., a human user behaviour), \textit{background} (e.g., some passive and automatic communications), \textit{botnet} and \textit{CnC}. Hence \textit{the benign part} of \dataset{CTU13} (in its 2 classes) can be used for the considered 5G slicing application proposed by the state-of-the-art. In this CS, we thus only consider the benign traffic of \dataset{CTU}, and exclude any malicious communication.
The vast majority of NetFlows (over 95\%) are for background.

We consider an attacker that controls 6 UEs (i.e., hosts) that perform both \textit{normal} and \textit{background} communications. Multiple PCAP traces are contained in \dataset{CTU13}, each captured at a different date. For our experiments, we use 8 out of the 12 provided traces, choosing those that contain a significant amount of traffic generated by the 6 UEs `owned' by the attacker. We consider each trace independently, i.e., we use each trace to: extract its NetFlows, label them, and devise a specific ML classifier trained and tested on the same trace. Such workflow allows to assess the effects of attacks in different circumstances, as each trace can be considered as a separate scenario.

For this CS, we cannot directly operate on the provided NetFlows because they are not raw-data, and are not valid for RsP. We are forced to operate on the PCAP traces.
Each PCAP trace of \dataset{CTU13} is used to generate NetFlows exactly as done by \dataset{CTU13} creators. Next, we label the generated NetFlows by using the exact procedure described in the ground truth information; we verify the correctness of our labelling scheme by cross-checking our labels with the original NetFlows in \dataset{CTU13}, \textit{and we obtain an exact matching}. 
We preprocess our NetFlows by removing missing values and computing additional features related to the \textit{IP addresses} and \textit{Ports}. We cannot use the exact IP addresses to perform the classification, as they would lead to overfitting. Therefore, we differentiate between \textit{internal} and \textit{external} IPs, whereas the Ports are categorized on the basis of the IANA guidelines (as  done in~\cite{apruzzese2020deep}). Hence, each NetFlow is described by the set of features $F$ in Table~\ref{tab:ctu_features}.
\begin{table}[htbp]
\centering
\caption{Features $F$ of \csctu\ . The affected $\overbar{F}$ are in gray.}
    \resizebox{0.6\columnwidth}{!}{
        \begin{tabular}{|c|c|c|}
        \hline
        \textbf{\#} & \textbf{Feature name} & \textbf{Type}  \\ \hline \hline
        1,2 & \textit{Src/Dst IP type} & Bool \\ \hline
        3,4 & \textit{Src/Dst port type} & Cat \\ \hline
        5 & \textit{Flow Direction} & Bool \\ \hline
        6 & \textit{Connection state} & Cat \\ \hline
        \rowcolor{gray!45}7 & \textit{Duration} (Dur) & Num \\ \hline
        8,9 & \textit{Src/Dst ToS} & Num \\ \hline
        \rowcolor{gray!45}10 & \textit{SrcBytes} (SrcByt) & Num \\ \hline
        \rowcolor{gray!45}11 & \textit{DstBytes} (DstByt) & Num \\ \hline
        \rowcolor{gray!25}12 & \textit{TotBytes} (Byt) & Num \\ \hline
        \rowcolor{gray!25}13 & \textit{TotPkts} (Pkt) & Num \\ \hline
        \end{tabular}
    }
\label{tab:ctu_features}
\end{table}

We partition our NetFlows in $T$ and $V$ with a 80:20 split (common in NetFlow analyses~\cite{apruzzese2020deep}). We considered many ML classifiers, but the \textit{RF} outperformed the others so we use this algorithm for our baselines.

To create the RsP, we take the raw PCAP trace and extract all raw traffic of the 6 controlled UEs of the attacker. We then consider every packet originating by such 6 UEs, and we increase its payload by appending small chunks [0-300] of random bytes; for TCP packets (which are stateful), we also ensure that the three-way handshake is carried out; we also ensure that the new packets will have the correct checksum; finally, we verify the integrity of the resulting PCAP trace and discard any inconsistent packets. This `myopic' PCAP trace is then subject to the same procedure as the original PCAP trace: we extract, label and preprocess the NetFlows. In Table~\ref{tab:ctu_features}, we denote with a light-gray background the features `consciously attacked' representing $\mathcal{\overbar{F}}$, and with a darker background the features $\overbar{F}$ that are influenced as a consequence of $\mathcal{\overbar{F}}$.

For the attacks at inference stage, we submit the myopic NetFlows to the trained \textit{RF}. For attacks at training-stage, we randomly take a portion (e.g., 25\%) of the NetFlows in $T$ that involve the UEs controlled by the attacker, and replace it with as many myopic NetFlows. 
We then retrain the \textit{RF} using such `poisoned' $T$ and test it again on $V$ to assess the effects on the whole environment.

All these operations are performed for each of the 8 PCAP traces. The results shown in Fig.~\ref{fig:results_ctu} report the average performance of all these experiments.

%% file: sections/app_CS/app_elasticmon.tex
\subsection{\cselasticmon\ : dataset, preprocessing, RsP}
\label{app:elasticmon}

The \dataset{ElasticMon} dataset is created by leveraging the Mosaic5G FlexRAN controller\footnote{\url{https://mosaic5g.io/apidocs/flexran/flexran_spec_v2.2.3.html}}. The samples capture information related to the MAC, RRC and PDCP. \dataset{ElasticMon} is provided either as raw-data or as preprocessed features. We use the raw-data as basis, to which we apply the same preprocessing as in~\cite{vasilakos2020integrated}, and use the same 90:10 split for creating $T$ and $V$. After preprocessing, we obtain the set $F$ of 16 features used by the \textit{RF} regressor\footnote{To compute accuracy, we round each predicted CQI to the nearest integer.}, which we report in Table~\ref{tab:elasticmon_features}. Here, a light-gray background denotes the features known and targeted by the myopic attacker ($\overbar{\mathcal{F}}$); while a dark-gray background are the features $\overbar{F}$ altered as a consequence of $\overbar{\mathcal{F}}$. All these features are numerical. We recall the 4 attack scenarios: ${\mathcal{\overbar{F}}_1}$=(RSRP); ${\mathcal{\overbar{F}}_2}$=(\textit{Byt}); ${\mathcal{\overbar{F}}_3}$=(\textit{Pkt}); ${\mathcal{\overbar{F}}_4}$=(\textit{Pkt,Byt}).

\begin{table}[!htbp]
  \centering
    \caption{Features $F$ of \cselasticmon\ . The affected $\overbar{F}$ are in gray.}
    \label{tab:elasticmon_features}
    \resizebox{0.5\columnwidth}{!}{
        \begin{tabular}{|c|c||c|c|}
                \hline
                \textbf{\#} & \textbf{Name} & \textbf{\#} & \textbf{Name} \\ \hline \hline
                \cellcolor{gray!45}1 & \cellcolor{gray!45}RSRQ & 9 & totTbsUL\\ \hline
                \cellcolor{gray!25}2 & \cellcolor{gray!25}RSRP & 10 & pktRxSn \\ \hline
                3 & PHR & \cellcolor{gray!25}11 & \cellcolor{gray!25}pktRx \\ \hline
                4 & totPrbDL & \cellcolor{gray!25}12 & \cellcolor{gray!25}pktRxByt \\ \hline
                5 & totPduDL & 13 & pktTxSn \\ \hline
                6 & totTbsDL & \cellcolor{gray!45}14 & \cellcolor{gray!45}pktRxAiat \\ \hline
                7 & totPrbUL & 15 & pktTxAiat \\ \hline
                8 & totPduUL & 16 & SFN \\ \hline
            \end{tabular}
    }
    
\end{table}

In particular, we observe that the RSRP is linked to the RSRQ, so we must also change the latter when doing the RsP for $\overbar{\mathcal{F}}_1$. Similarly, increasing the packets (\textit{pktRx}) will also change their inter arrival time (\textit{pktRxAiat}), which must be updated when applying the RsP for $\overbar{\mathcal{F}}_2$ and $\overbar{\mathcal{F}}_4$.

When determining the intensity of each RsP, we proceed as follows. For $\overbar{\mathcal{F}}_1$, we change the RSRP randomly with another value in \dataset{ElasticMon}; we do this 7 times. For the remaining attacks (which target the \textit{pktRx} and \textit{pktRxByt}), we increase the corresponding values of each sample as a function of their standard deviation in 7 intensity steps $\mathcal{I}$. Specifically, we use:
\vspace{-1em}
\begin{align*}
\label{eq:intensity_elasticmon}
    \text{$\mathcal{I}$} \in \big(\{0.1, 0.2, 0.5, 1, 2, 5, 10\} \times std_f\big)
\end{align*}
where ${std}_f$ is the standard deviation of the feature $f$ in \dataset{ElasticMon}.
Let us explain our workflow by using $\overbar{\mathcal{F}}_2$=\textit{pktRx} as an example. We take the raw-data in \dataset{ElasticMon}, from which we compute the standard deviation of the \textit{pktRx} metric $std_{pktRx}$; next, for each intensity value $i \in \mathcal{I}$, we multiply $std_{pktRx}$ by the considered intensity $i$; then, we add this value to the \textit{pktRx} of each sample in the \dataset{ElasticMon}. This creates multiple (7) sets of adversarial raw-data, each associated to a given intensity value $i \in \mathcal{I}$ targeting the features of $\overbar{\mathcal{F}}_2$. We pre-process each of these sets with the same procedure as in~\cite{vasilakos2020integrated}, where we also update the derived \textit{pktRxAiat} feature. We then submit these myopic sets to the baseline \textit{RF} regressor at inference stage.

We now explain our application of the two countermeasures.
We apply \textit{feature removal} 4 times, each time by re-training the baseline \textit{RF} on the same $T$ but without considering the features influenced by each of the 4 myopic attacks. 
We also apply \textit{adversarial training} 4 times, by isolating a small portion (5\%) of $T$, denoted $T_{aug}$, on which we apply the RsP of each specific myopic scenario at all intensity levels. We then re-train the baseline \textit{RF} on the original $T$ as well as all the `perturbed' variants of $T_{aug}$.

\begin{figure}[!htbp]
    \centering
    \includegraphics[width=0.95\columnwidth]{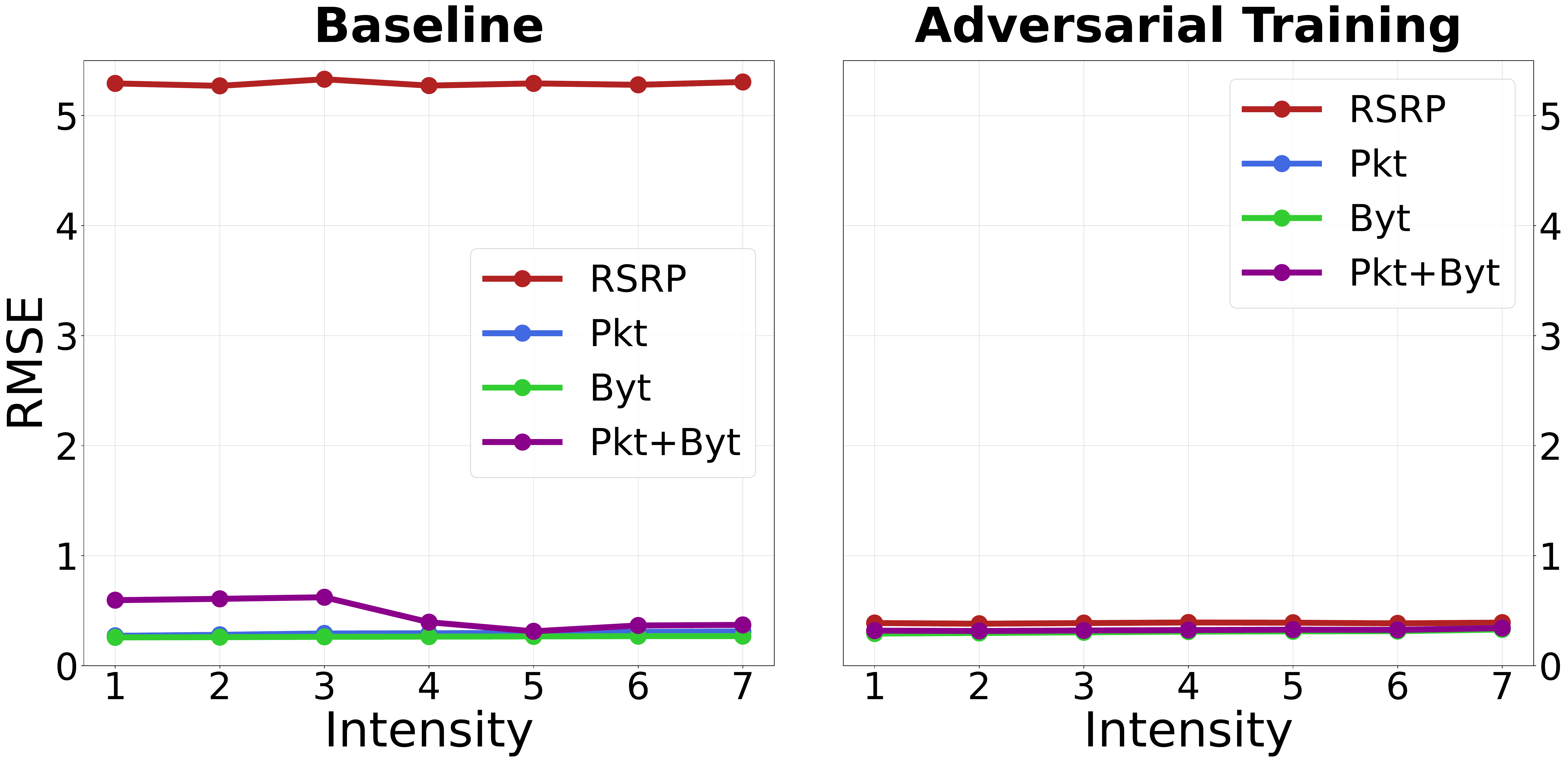}
    \caption{\cselasticmon\ : Attacks and Defense (Baseline \textit{RMSE}=$0.25$)}
    \label{fig:results_elasticmon_rmse}
\end{figure}

For completeness, we present in Fig.~\ref{fig:results_elasticmon_rmse} the efficacy of our attacks as measured via \textit{RMSE} (lower is better) instead of accuracy as in Fig.~\ref{fig:results_elasticmon}, shown in the main paper.

%% file: sections/app_CS/app_irish.tex
\subsection{\csirish{}: time series}
\label{app:irish}

\begin{figure}[!htbp]
    \centering
    \begin{subfigure}[t]{0.24\textwidth}
        \centering
        
        \includegraphics[width=\columnwidth]{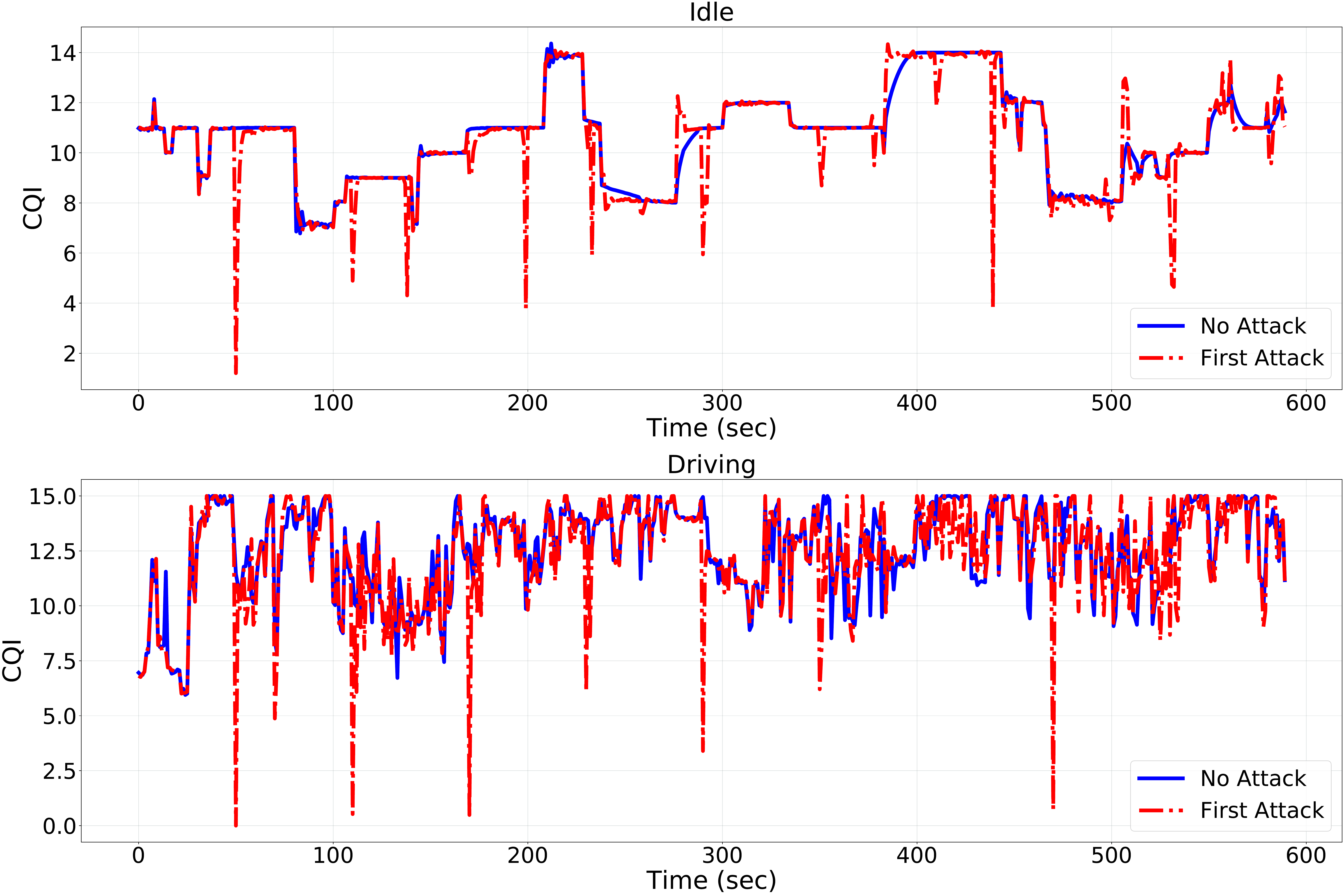}
         \caption{\footnotesize{1st attack: CQI=0 every 60s.}}
         \label{sfig:irish_a1}
    \end{subfigure}%
    ~ 
    \begin{subfigure}[t]{0.24\textwidth}
        \centering
        
        \includegraphics[width=\columnwidth]{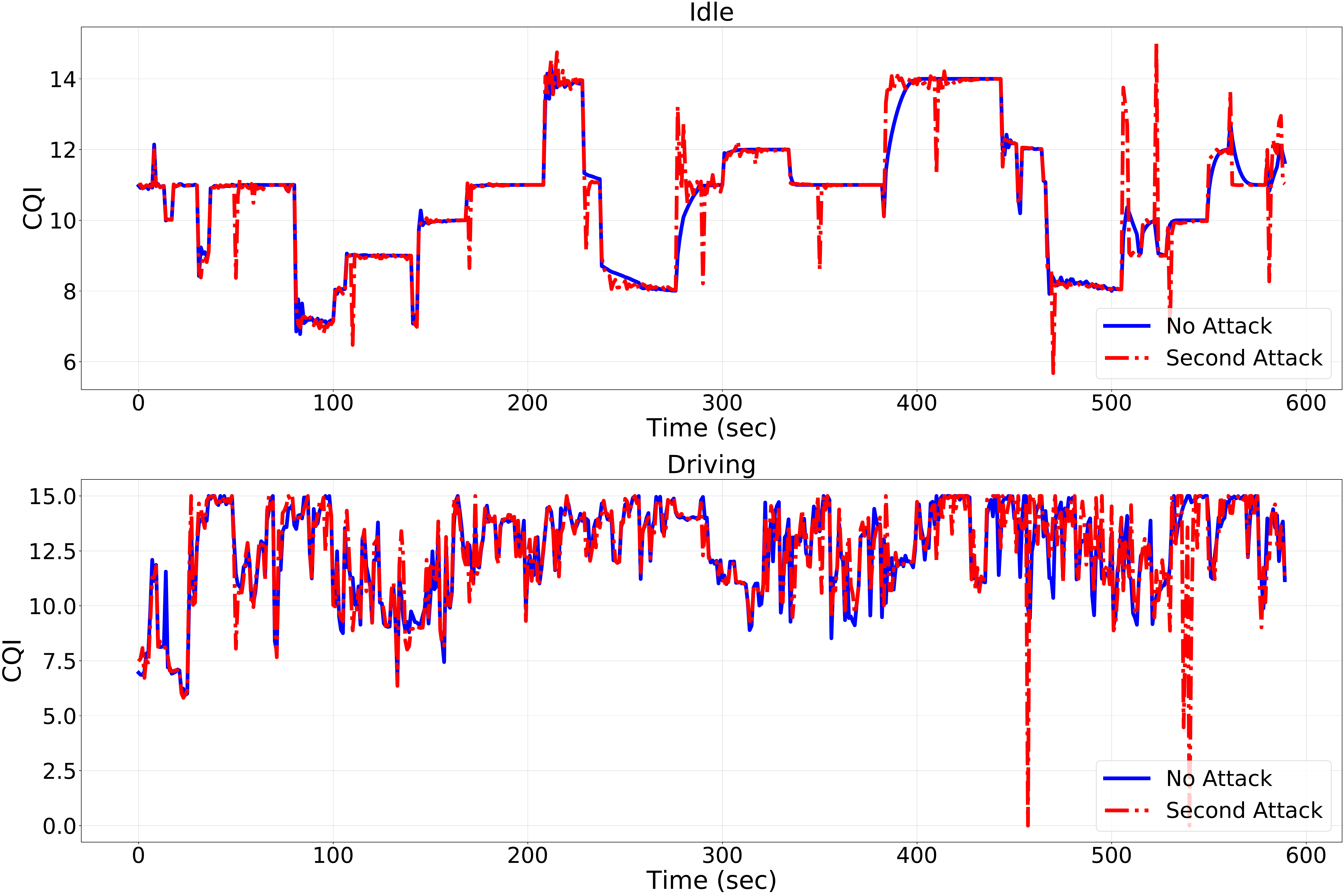}
         \caption{\footnotesize{2nd attack: CQI$\pm$3 every 60s.}}
         \label{sfig:irish_a2}
    \end{subfigure}
    \caption{\csirish\ : predicted CQI with and without myopic examples.}
    \label{fig:irish_timeseries}
\end{figure}

We report in Figs.~\ref{fig:irish_timeseries} the time series of the predicted CQI. The blue lines correspond to predictions on clean data, while red lines correspond to adversarial behavior of the two attacks considered in the paper. The time series at the top (bottom) of Figs.~\ref{fig:irish_timeseries} correspond to the `static' (`driving') mobility pattern.

%% file: sections/app_CS/app_rml.tex
\subsection{\csrml: Dataset, RsP, related work}
\label{app:rml}

We considered many `shallow' ML classifiers, but the \textit{RF} outperformed the others so we use this algorithm as shallow-learning baseline, whereas the deep-learning baseline uses the same \textit{DN} architecture as prior work.
To devise our baselines, we take the \dataset{RML2016.10a} dataset, filter the modulations pertaining to 5G (as done in~\cite{usama2021examining}, specifically: BPSK, QPSK, 8PSK, CPFSK, GFSK, WBFM), and partition it with the same 50:50 split as in~\cite{kim2020over} into $T$ and $V$. We use the exact \textit{DN} architecture as in previous work~\cite{o2016radio, kim2020over}, while the \textit{RF} is fine-tuned to ensure optimal classification performance.

For each of the 4 considered attack scenarios, we apply the RsP in the same way as in \cselasticmon\ ~(described in Appendix~\ref{app:elasticmon}).
We first determine the measurements to perturb: randomly chosen (either 25, 50 or 100 for $\mathcal{\overbar{F}}_1$, $\mathcal{\overbar{F}}_2$, $\mathcal{\overbar{F}}_3$, respectively) or by using the top-25 most significant in the worst-case ($\mathcal{\overbar{F}}_4$).
Then, we increase the corresponding measurements of each sample as a function of their standard deviation, in 7 intensity steps as done in \cselasticmon\. We ensure that all such increments fall within acceptable ranges. Finally, we note that I/Q measurements are orthogonal~\cite{wen2017enhanced}, hence $\overbar{F}$=$\mathcal{\overbar{F}}$.

Because the attacks in $\mathcal{\overbar{F}}_{1-3}$ involve a lot of randomness, we repeat these attacks 20 times (each time by choosing different measurements). In our results (in Fig.~\ref{sfig:rml_myopic}) we report the average values and their standard deviation (as a black vertical line on each marker). For the attacks of related work~\cite{usama2021examining, kim2020over}, we report the values directly as shown in the corresponding paper because they target a very similar system: the only difference is that~\cite{kim2020over} does not remove the analog modulations (but their \textit{DN} obtains the same baseline performance as ours), whereas~\cite{usama2021examining} uses only the samples with SNR=10db to train and test their \textit{DN}, which explains the slight discrepancy in baseline performance---reported as a black line in Fig.~\ref{sfig:rml_related}.

%% file: sections/app_CS/app_deepslice.tex
\subsection{\csdeepslice\ : details and insider myopic attacks}
\label{app:deepslice}
The \dataset{DeepSlice} dataset uses the 5G specifications to describe KPI representing the main 5G use-cases (eMBB, MMTC, URLLC), each denoting a slice and, hence, a label.

We do not perform any pre-processing to \dataset{DeepSlice}. We report in Table~\ref{tab:deepslice_features} the feature set $F$ describing each sample and analyzed by the baseline \textit{DN}. We partition the dataset by using the same split as~\cite{thantharate2019deepslice} of 90:10 for $T$ and $V$, matching their perfect accuracy: \textit{Acc}=$1.00$. A true myopic attacker can only affect the \textit{day} and \textit{hour}, but such attempts are not successful against the \textit{DN}, as shown in §\ref{ssec:deepslice}.

\textbf{`Insider' myopic Attacks.}
An `insider' myopic attacker that can operate from within the 5G NI and tamper with the gNB configurations. She can thus affect the \textit{PktLossRate} or the \textit{PktDelayBudget} associated to each slice, thus ${\mathcal{\overbar{F}}}$=(\textit{PktDelayBudget}, \textit{PktLossRate}). We design 3 attack scenarios: $\mathcal{\overbar{F}}_1$=(\textit{PktDelayBudget}), $\mathcal{\overbar{F}}_2$=(\textit{PktLossRate}), 
$\mathcal{\overbar{F}}_3$=(\textit{PktDelayBudget}, \textit{PktLossRate}); in all these cases, $\overbar{F}$=$\mathcal{\overbar{F}}$, shown in gray in Table~\ref{tab:deepslice_features}. 

We craft the RsP in the same way as described in Appendix~\ref{app:elasticmon}. All the affected features are independent.

We counter such attacks with \textit{adversarial training} and \textit{feature removal}, applied in the same way as in \cselasticmon\ ~(§\ref{ssec:elasticmon}).

\begin{table}[htbp]
\centering
\caption{$F$ of \csdeepslice\ . Gray rows denote the insider $\overbar{F}$.}
    \resizebox{0.5\columnwidth}{!}{
        \begin{tabular}{|c|c|c|}
        \hline
        \textbf{\#} & \textbf{Feature name} & \textbf{Type}  \\ \hline \hline
        1 & UseCase & Cat \\ \hline
        2 & UEcategory & Cat \\ \hline
        3 & Technology & Cat \\ \hline
        4 & Day & Num \\ \hline
        5 & Hour & Num \\ \hline
        6 & GuaranteedBitRate & Bool \\\hline
        \rowcolor{gray!45} 7 & PacketLossRate & Num \\\hline
        \rowcolor{gray!45} 8 & PacketDelayBudget & Num \\\hline
        
 \hline
        \end{tabular}
    }
\label{tab:deepslice_features}
\end{table}

\begin{figure}[!htbp]
    \centering
    \includegraphics[width=0.5\columnwidth]{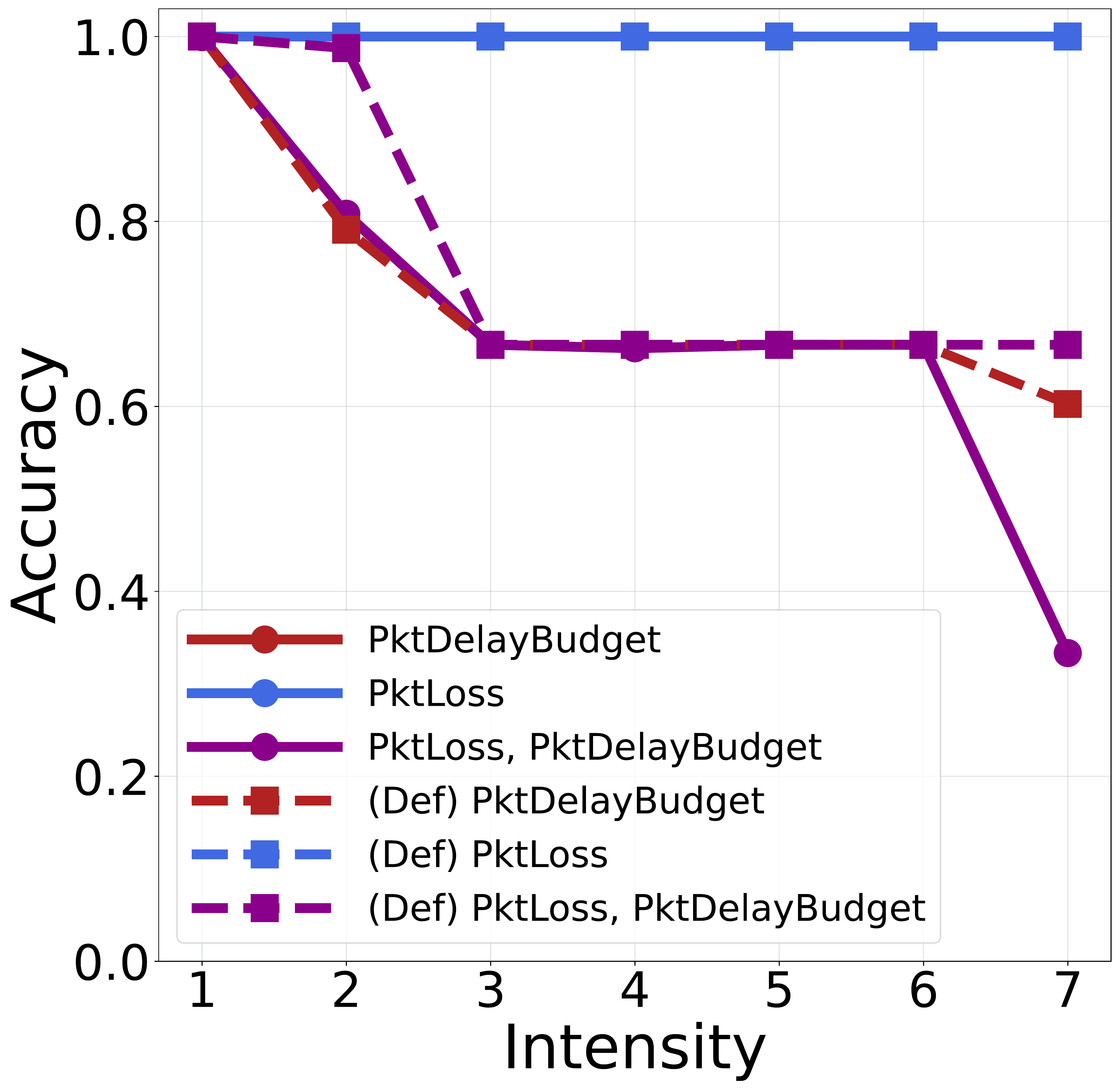}
    \caption{\csdeepslice\ : Attack and Defense (Base: \textit{Acc}=$1.00$)}
    \label{fig:results_deepslice}
\end{figure}

\textbf{Results.} 
We report the results of these insider attacks in Fig.~\ref{fig:results_deepslice}, reporting accuracy as a function of RsP intensity. We use full lines for attacks against the baseline \textit{DN}, and dotted lines for attacks against the \textit{DN} hardened via adversarial training; as in CS3, feature removal always defused the attacks. The tradeoff of the countermeasures is shown in Table~\ref{tab:tradeoff_deepslice}. We observe that $\mathcal{\overbar{F}}_2$ has no effect, but $\mathcal{\overbar{F}}_1$ and $\mathcal{\overbar{F}}_3$ can decrease the accuracy even at low intensities. Feature removal induces an unfavorable tradeoff; in contrast, adversarial training preserves baseline performance but its protection is low.

\begin{table}[!htbp]
    \centering
    \caption{\csdeepslice\ -insider: $\mathbb{T}$ (lower is better, $\mathbb{T}$=$1$ is no change).}
    \label{tab:tradeoff_deepslice}
    \resizebox{0.5\columnwidth}{!}{
        \begin{tabular}{c?c|c|c}
             \toprule
             
             \textbf{Defense} & ${\mathcal{\overbar{F}}_1}$ & ${\mathcal{\overbar{F}}_2}$ & ${\mathcal{\overbar{F}}_3}$ \\
             
             \midrule
             Adv. Tra. & $1.00$ & $1.00$ & $1.00$\\
             Fea. Rem. & \cellcolor{red!10}$1.50$ & $1.00$ & \cellcolor{red!10}$1.50$\\
             
             \bottomrule
        \end{tabular}
    }
\end{table}